\documentclass[pre,twocolumn,amsmath,amssymb,aps]{revtex4}

\usepackage{graphicx, epstopdf, subfigure}
\usepackage{dcolumn}
\usepackage{bm}

\newcommand{\change}[1]{{\bf #1}}

\graphicspath{{data/}}
\begin{document}

\preprint{APS/123-QED}

\title{Rare-Event Properties of the Nagel-Schreckenberg Model}

\author{Wiebke Staffeldt}
 \affiliation{Institut f\"ur Physik, Unversit\"at Oldenburg,
26111 Oldenburg, Germany}
\author{Alexander K. Hartmann}%
 \email{a.hartmann@uni-oldenburg.de}
 \affiliation{Institut f\"ur Physik, Unversit\"at Oldenburg,
26111 Oldenburg, Germany}

\date{\today}

\begin{abstract}
We have studied the distribution of traffic flow $q$ for the 
Nagel-Schreckenberg model by computer simulations. 
We applied a
large-deviation approach, which allowed us to obtain the
distribution $P(q)$ over more than one hundred decades in probability,
down to probabilities like $10^{-140}$. This allowed us to
characterize the flow distribution over a large range of the support
and identify the characteristics of rare and even very rare
traffic situations. We observe a change of the distribution shape
when increasing the density of cars from the free flow to the
congestion phase. Furthermore, we characterize typical and rare
traffic situations by measuring correlations of $q$ to other
quantities like density of standing cars or number and size of traffic jams.
\end{abstract}

\maketitle


\section{\label{sec:intro}Introduction}

The Nagel-Schreckenberg model for traffic flow is
a cellular automaton model introduced in 1992 \cite{NaSch}. 
It is a very simple yet fundamental
model not only for traffic but also for general transport
phenomena and the occurrence of slow (glassy) dynamics \cite{deWijn2012}.
This model has started in the statistical physics
community a very active field of research on traffic \cite{SchadRev,HelbRev}.
To mimic human behavior in the model, 
which leads to fluctuations and in turn to the occurrence of spontaneous
traffic jams, a
stochastic component responsible for random speed reduction
was introduced. The model undergoes a change of the behavior
from free flow to congested traffic when the density of cars increases
\cite{NaSch}, reminding of a phase transition.
 In the past years, a lot of research concerning this
model was on the question, whether this transition is an example for a
real non-equilibrium phase transition \cite{SchadRev, Eisen, rotors,
comment}. 

So far in  literature, the model was investigated,
let it be analytically or numerically, regarding its \emph{typical} behavior.
Usually, the fundamental diagram, which shows the mean value of
traffic flow $q$ as a function of the density of cars $\rho$ is
studied. It shows a maximum at a density $\rho_{\max}$
\cite{NaSch}, which signifies the change from free flow
to congested behavior.  
However, investigating typical behaviour
 only allows, from a fundamental point of view,
 to explore the model with a rather narrow focus. 

Nevertheless, when describing any stochastic 
 model as comprehensively as possible,
one aims at obtaining the probability distributions of the quantities
of interest over a large range of the support. This means
one is interested in understanding the model beyond its typical
properties down to \emph{rare}
events that only occur with very small probabilities, 
also called \emph{large deviations} 
\cite{denHollander2000,touchette2009,dembo2010,touchette2011}. 

Furthermore, for traffic models this is
also of high practical interest, because even minor 
fluctuations can, with a small probability,
lead to severe consequences like large traffic jams.  Thus also
rare events should be studied to understand them better and possibly 
help to plan traffic system such that costly
 traffic jams due to rare-event fluctuations 
are better avoided in real situations. Therefore, a large-deviation approach
\cite{work} based on a Monte Carlo (MC) simulation 
is applied here by which the measure $q$ can also be
obtained in atypical regions. Thus, it enables one to obtain the
distribution of the traffic flow $q$ over the full support, or at
least a much larger range of values than standard approaches. 
 We are able to resolve
the distribution of $q$ in some cases over more than one 100
orders of magnitude, down to probability densities as small as $10^{-140}$.

It is the aim of this work to investigate whether the
shape of the distribution of the traffic flow $q$, 
in particular its tail
properties,  changes at the transition
from free to congested flow.
 The analysis of the distributions is done by approximating
the tails with a fit function, and then studying the fit parameters as
a function of the density of cars. 
 In addition, the large-deviation approach enables us to
investigate correlations between the traffic flow $q$ and other
measures like the average jam size $S_{\rm av}$ or the density of jammed
cars $\rho_{\text{jammed cars}}$ over a larger range of values. 
This allows us to characterize the
atypical states of the system.

The paper is organized as follows. Next, we define the model
and explain the numerical approaches we applied. 
In section \ref{sec:distr} the results for the distributions
are presented and
compared to results obtained from previous work \cite{NaSch,Krug,
relax}. 
 The results concerning correlations of $q$ with other values
of interest are displayed
in section \ref{sec:corr}. We finish by a discussion and outlook in section
\ref{sec:discussion}.

\section{\label{sec:method}Model and Methods}

The Nagel-Schreckenberg model, as studied here by computer simulations
\cite{practical_guide2015}, is defined on an
one-dimensional lattice with $L$ sites and periodic boundary
conditions. Each car $i$ has a position
$x_i \in \{0, 1, ..., L-1\}$ and a velocity $v_i \in \{0, 1, ...,
v_{\max}\}$ with $v_{\max}$ being the speed limit. With the total number
of cars $N$, the density of cars is
\begin{equation}
\rho = \frac{N}{L}
\end{equation} and
the traffic flow $q$ (per lattice site)
is given by  
\begin{equation}
q = \frac{1}{L}\sum_{i=0}^{N-1} v_i\,.
\end{equation}
 
The dynamics of the system is described by the following
update-rules, that are applied one after the other, in parallel for all vehicles
\cite{NaSch}:
\begin{enumerate}
\item {\bf Acceleration}:
if car $i$ has a velocity $v_i < v_{\max}$: 

\centerline{$v_i \rightarrow v_i+1$}

\item {\bf Slowing down} due to other vehicles, preventing accidents:
if the gap $d_i = x_{i+1}-x_i-1$ between a car $i$ and the preceding car $i+1$ 
is smaller than the velocity $v_i$ of car $i$:

\centerline{$v_i \rightarrow d_i$}

\item {\bf Randomization}:
with probability $p$, the velocity $v_i$ of car $i$ changes to:

\centerline{$v_i \rightarrow \max\left(v_i-1,0 \right)$}

\item {\bf Movement}:
The position $x_i$ of car $i$ is updated:

\centerline{$x_i \rightarrow x_i + v_i$}

\end{enumerate}

For our simulations, we have chosen  $v_{\max} = 5$ as in the original 
publication, for convenience and $p = 0.2$.
 The general results should not depend much on the
choice of these values. Starting with any initial configuration, simulating the
Nagel-Schreckenberg dynamics is straightforward and fast. To get a
description of any quantities of interest, in particular of the
traffic flow $q$, one has to measure the distribution of these
quantities, e.g., $P(q)$.  In a straight forward way, one could simulate
the system over long times and collect sample values in
histograms. This sampling according the natural probabilities we
call \emph{simple sampling} (sometimes also called importance sampling
in the literature).  The number $J$ of statistically independent
samples, typically $J\sim 10^5-10^9$,
 gives an indication of how well the distributions can be
sampled, the smallest probabilities which can be approximately
obtained are $O(J^{-1})$.  If one is interested in obtaining the
distributions over a substantial part of the support, one needs to
measure in regions where the probabilities are much smaller, like
$\sim 10^{-50}$. Here large-deviations algorithms can be used.

\begin{figure} [ht]
\begin{center}
\includegraphics[width = 0.32 \textwidth,  angle =270]{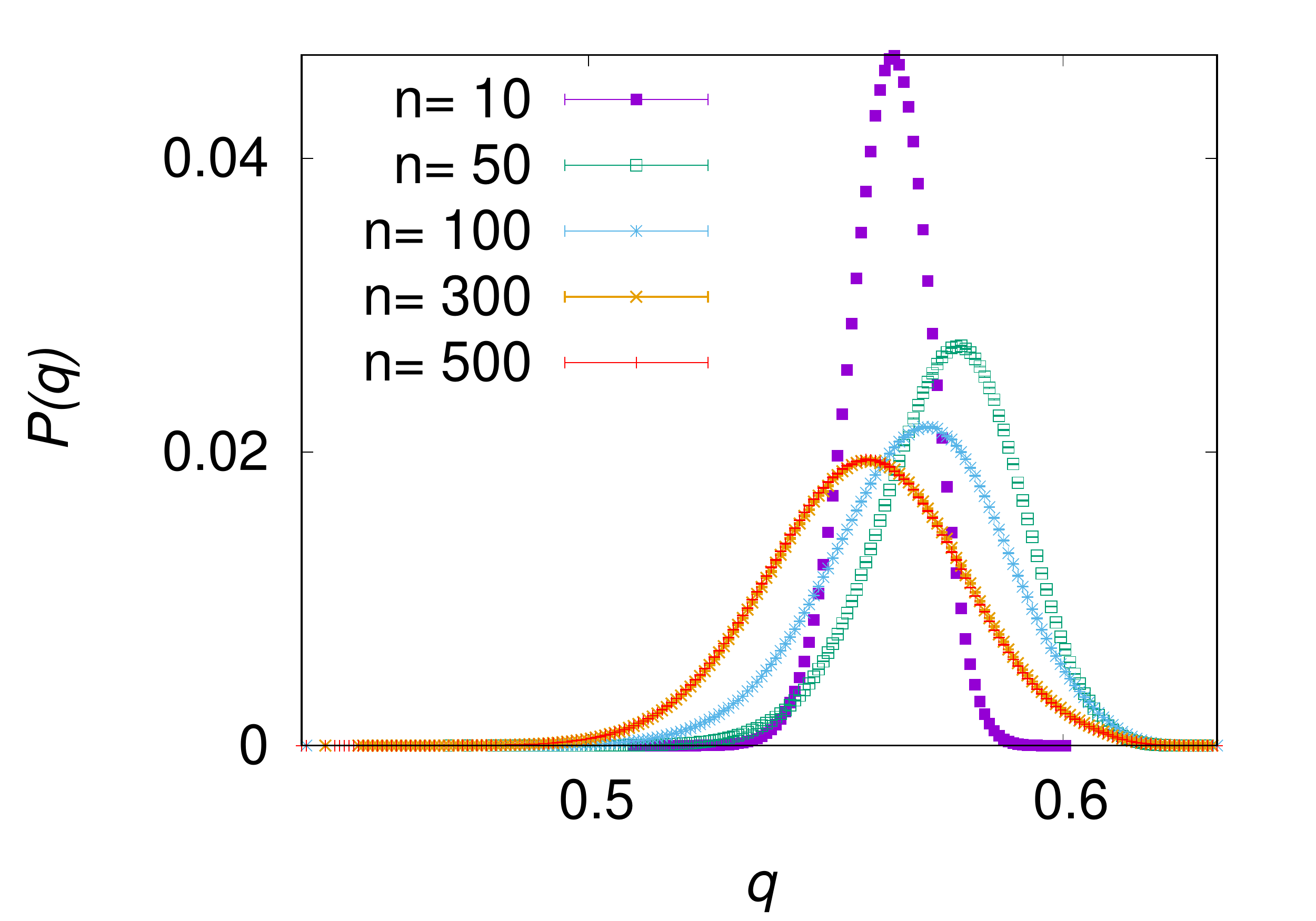} 
\caption{\label{fig:Dn} 
(color online) Distribution of the traffic flow $q$ for a system of size 
$L=1000$ and $\rho = 0.13$ where repeatedly, 
starting  always with the same (typical) configuration $\underline{y}^{(0)}$, 
statistically independent histories of $n$ steps were
performed and then $q$ was measured each time after 
these $n$ steps. The distributions  
are shown for various number $n$ of steps.}

\end{center}
 \end{figure}

For self-containedness, and to include the adaptations made for the
present model, we include a brief description of 
the large-deviation approach used here, which was basically introduced 
previously \cite{work} for the case of the distribution of work for an
Ising model. For the present study  we consider a sequence 
$\underline{Y}=(\underline{y}^{(0)},\underline{y}^{(1)}, \ldots 
\underline{y}^{(n)})$ of traffic configurations ($\underline{y}$
consisting each of positions 
$\{x_i\}$ and velocities $\{v_i\}$), where the transitions
$\underline{y}^{(t)}\to \underline{y}^{(t+1)}$  are according to the 
Nagel-Schreckenberg model presented above. We call $\underline{Y}$ a \emph{history}.
We assume that the initial
configuration $\underline{y}^{(0)}$ 
is in a steady state, which can be obtained by performing a long-enough
simulation starting from any configuration at a given density.  Furthermore,
we assume that  $n$ is sufficiently
large such that $\underline{y}^{(n)}$ is statistically independent
of $\underline{y}^{(0)}$. Thus, when measuring the traffic flow
$q=q(\underline{Y})$ (only) at the final configuration 
$\underline{y}^{(n)}$, one obtains a statistically independent
measurement of $q$. To obtain \change{$K$} independent values of $q$,
instead of performing one very long simulation with measuring every $n$ steps, 
one can therefore also perform
\change{$K$} independent runs each generating a history $\underline{Y}$ of 
length $n$, all
starting with the same initial configuration, but with statistically 
independent  evolutions. This scheme will it make possible to apply the 
large-deviation approach as explained below.
Since the value of $n$ will be chosen large enough, the fact
that $\underline{y}^{(0)}$ is fixed plays no role.
Now we discuss the choice of $n$.
Figure \ref{fig:Dn} shows the distributions $P_n^{(0)}(q)$ 
for a system of size $L=1000$
and $\rho = 0.13$ after repeatedly evolving the system for a $n$ number of
steps  starting from the same steady state configuration $\underline{y}^{(0)}$. 
The data
shown in Figure \ref{fig:Dn} was obtained via simple sampling.
For a small number $n$ of time steps, the histograms strongly depends on the
fixed initial configuration, i.e., the histogram is very narrow (for $n=0$
it is a delta peak at $q(\underline{y}^{(0)})$).
 When increasing $n$, the histograms change considerable, in particular they
become broader.
This change with $n$ becomes weaker when increasing $n$ even more, 
representing increasing statistical independence of the measured values $q$
from the traffic flow of the initial configuration. 
The distributions are almost identical for $n = 300$ and $n =
500$. Therefore, $n = 300$ should approximately correspond to the
correlation time of the system since this seems to be the amount of
time steps it takes the system to forget its initial configuration. 
Thus, $n = 300$ is used throughout this work. 

Now, the main idea to access the low-probability tail is 
not to generate a sample with histories $\underline{Y}$,
where each state occurs with its natural probability
$R(\underline{Y})$ \cite{graph}. Instead one 
samples according to a biased
distribution \cite{hammersley1956,bucklew2004}, 
where each history $\underline{Y}$ occurs with a
probability taken as 
\begin{equation}
R_\theta(\underline{Y})  =
\frac{1}{Z(\theta)}R(\underline{Y})e^{-q(\underline{Y})/\theta}
\label{eq:bias}
\end{equation}
 with
a normalization factor $Z(\theta)$ and an \emph{artificial temperature}
$\theta$, see Ref.~\cite{graph}. The bias factor $^{-q(\underline{Y})/\theta}$
allows to control the range of sampled configurations via 
choosing $\theta$ appropriately. In particular for infinite temperate
we recover simple sampling, i.e., we have  
$R_\infty(\underline{Y})  = R(\underline{Y})$.
When sampling configurations from 
$R_\theta(\underline{Y})$ and measuring $q=q(\underline{Y})$ one encounters
a biased distribution
$P_\theta(q)$ for the traffic flow $q$, depending on  temperature $\theta$,
which is given by \cite{align2002}

\begin{align}
P_\theta(q)& = \sum_{\underline{Y}} R_\theta(\underline{Y}) 
\delta_{q(\underline{Y}), q}\nonumber \\
&= \sum_{\underline{Y}}\frac{1}{Z(\theta)}R(\underline{Y})e^{-q(\underline{Y}) / \theta} \delta_{q(\underline{Y}), q} \nonumber \\
&=\frac{e^{-q / \theta}}{Z(\theta)} \sum_{\underline{Y}} R(\underline{Y})\delta_{q(\underline{Y}), q} \nonumber\\
&= \frac{e^{-q / \theta}}{Z(\theta)} P(q), \label{eq:P-theta-q}
\end{align}
with $P(q) = \sum_{\underline{Y}} R(\underline{Y})\delta_{q(\underline{Y}), q}$ 
being the actual target 
probability, i.e., the probability
 to measure a certain value of the traffic flow 
$q$ when each history $\underline{Y}$ occurs with its natural probability 
$R(\underline{Y})$ \cite{graph}.\\

To generate histories $\underline{Y}$
according to $R_\theta (\underline{Y})$, we used 
the standard Metropolis-Hastings
MC algorithm. Therefore, a Markov chain 
$\underline{Y}(0), \underline{Y}(1), \underline{Y}(2), \ldots$
of histories is created. This numbering of histories defines a second time
in our simulation approach,
the Markov-chain time $t_{\rm MC}$, not to be confused
with the above inbtroduced 
time $t$ of the Nagel-SChreckenberg traffic evolution, which is relevant within
each of the histories.
  In each step $t_{\rm MC}+1$, based on the current
history $\underline{Y}(t_{\rm MC})$, a
trial history $\underline{Y}_{\rm trial}$ with a traffic flow $q_{\rm trial} =
q(\underline{Y}_{\rm trial})$ is generated. 
This involves for each $\underline{Y}_{\rm trial}$ 
a complete simulation of $n$ steps according the Nagel-Schreckenberg model
and $q$ is measured for the final configuration.
The trial history is accepted to be the next
element of the Markov Chain, i.e., 
$\underline{Y}(t_{\rm MC}+1)=\underline{Y}_{\rm trial}$  
with the Metropolis acceptance probability
\cite{work}

\begin{equation}
\label{eq:MAP}
A(\underline{Y} \rightarrow \underline{Y}_{\rm trial}) = 
\text{min}\left(1,  e^{-[q(\underline{Y}_{\rm trial})-  
q(\underline{Y}(t_{\rm MC}))]/\theta}\right).
\end{equation}
If the trial history is not accepted, the next history will be
the current one, 
$\underline{Y}(t_{\rm MC}+1)=\underline{Y}(t_{\rm MC})$, as usually.

Next, we explain how we set up the simulation of a history, to
incorporate it into the Markov chain Monte Carlo approach.  In the randomization
step (number 3), random numbers are necessary to determine the braking 
of cars. In every iteration, each car needs exactly one random number to
decide whether to randomly decrease its velocity by one. 
In a typical implementation, one would call a pseudo random number
generator function each time a random number is needed.
In the large-deviation approach from \cite{work}, 
these random numbers are computed before
performing the actual simulation \cite{crooks2001} 
and stored in a vector $\underline{\xi}$. Whenever a random number is needed in the
simulation, it is taken from $\underline{\xi}$. This approach is just
a small, but necessary, technical adjustment, not at all altering
the simulation result. Thus,
the value of the measure, the traffic flow $q$, depends deterministically
on this vector
$\underline{\xi}$, since all randomness is subsumed here.
 Therefore, a Markov chain of histories $\underline{Y}(t_{\rm MC})$,
given the initial configuration $\underline{y}^{(0)}$, is
completely  determined by a Markov chain 
of vectors $\underline{\xi}^{(t_{\rm MC})}$ ($t_{\rm MC}=0,1,2,\ldots$)
because 
each history (at Monte Carlo time $t_ {\rm MC}$) will start
at the same configuration $\underline{y}^{(0)}$ 
and evolve for $n$ steps according $\underline{\xi}^{(t_{\rm MC})}$.
Since each of the $N$ cars needs at each traffic time step $t$ 
exactly one random
number, the vectors $\underline{\xi}$ covering all $n$ traffic time steps has size 
$n\times N$.
The observable of interest is the traffic
flow $q$ measured in the last time step, i.e., for $\underline{y}^{(n)}$.

To generate with the MC simulation a trial history
$\underline{Y}_{\rm trial}$, correspondingly  a trial vector
$\underline{\xi}_{\rm trial}$ is constructed. We perform
only small changes to obtain $\underline{\xi}_{\rm trial}$, thus
we change only a small number of
randomly selected entries of the current vector $\underline{\xi}(t_{\rm MC})$ 
\cite{work}. This is a local change to the current history, similar
to a standard local spin-flip when simulating magnetic systems using
Monte Carlo approaches. The amount of changed
entries is variable. We have chosen it  in a way that for each temperature 
$\theta$ approximately 
half of the trial histories get accepted \cite{work}, as a rule of thumb.

\begin{figure} [ht]
\begin{center}
\includegraphics[width = 0.5 \textwidth]{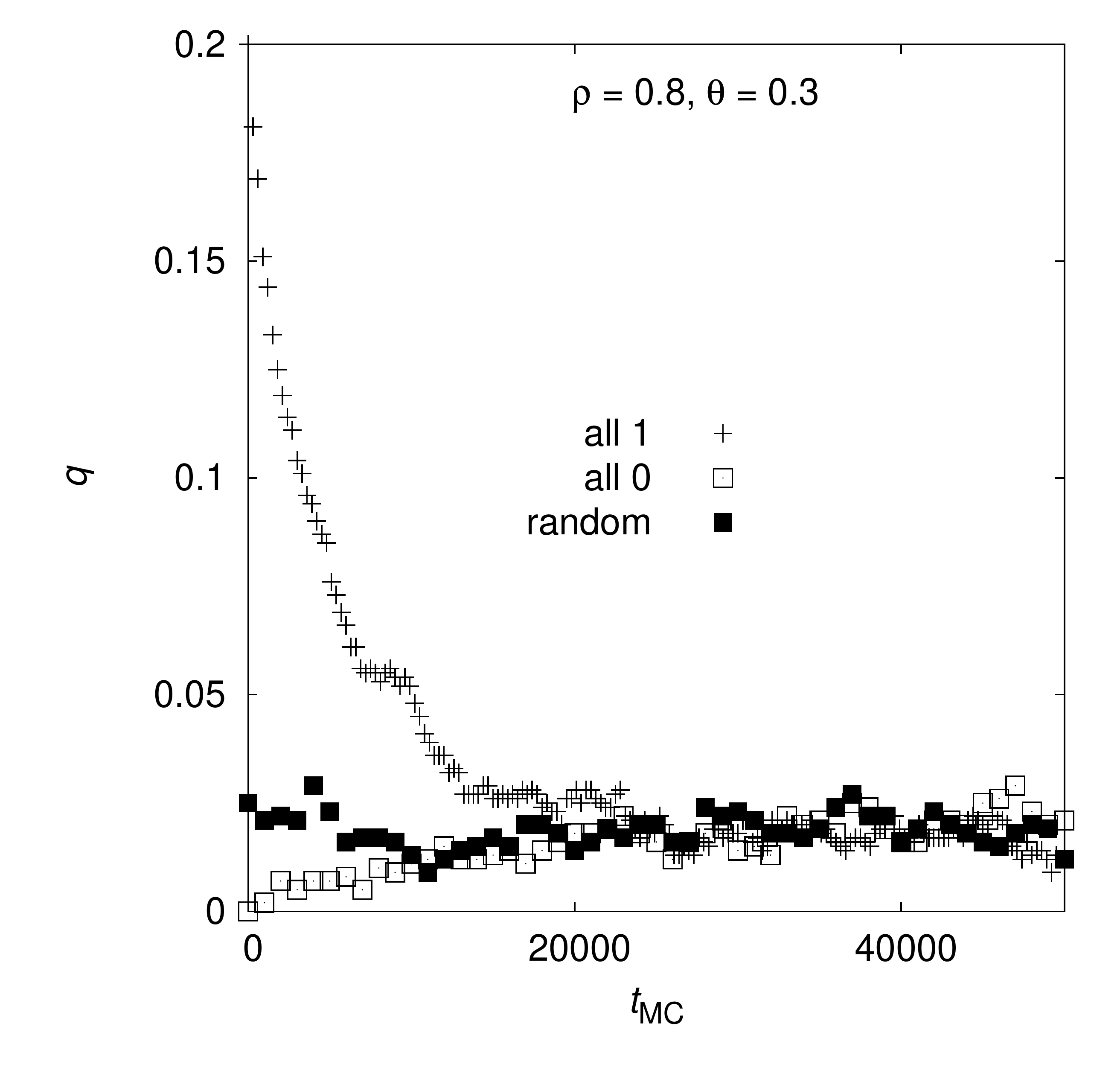} 
\caption{\label{fig:equi} Equilibration of the traffic flow $q$ as a function of the 
Monte Carlo time $t_{MC}$ for a system of size $L=1000$ 
and $\rho = 0.8$ for an artificial temperature $\theta = 0.3$.}
\end{center}
 \end{figure}

To ensure that the MC Simulation has equilibrated, the system
is prepared with different initial histories $\underline{Y}(0)$.  
Technically, this is
achieved by initializing the vector
$\underline{\xi}(0)$ with random numbers with either a) all entries
set to one, or b) with all entries set to zero, or c) 
with entries drawn from a uniform distribution in [0,1].
 Figure \ref{fig:equi} shows the measured values of the traffic flow
$q$ (at the final configuration of each history, as always here)
as a function of the Monte Carlo time $t_{MC}$ for a system with a
very high density $\rho = 0.8$ using different initializations. 
All curves reach, within fluctuations, the same value after
a sufficient number of MC steps,
 proving the equilibration of the MC simulations.
The
curve corresponding to the initialization with all entries set to zero
(which for our implementation 
means for the \emph{initial} history 
the cars will always break for all traffic time steps, i.e., never accelerate) 
reaches equilibrium from below, whereas the curve corresponding to the
initialization with all entries set to one (no spontaneous breaking initially) 
reaches the curve from above. Also for other values of $\rho$ and $\theta$
we have verified equilibration in a corresponding way.

After equilibration, the correct distribution $P(q)$ 
can be obtained from the measured biased distributions $P_\theta(q)$ 
(approximated by histograms)
according to Eq.~(\ref{eq:P-theta-q}) by \cite{work}
 \begin{equation}
 P(q) = e^{q / \theta}Z(\theta) P_\theta(q).
 \end{equation}
One has to perform simulations for several suitably chosen 
values $\theta_1,\theta_2,\ldots, \theta_K$, each
focusing the simulation to a certain range of values of $q$. One
can reconstruct $P(q)$ by combination of all results.
 The normalization constants $Z(\theta_k)$ can be calculated \cite{align2002}
by a global normalization of $P(q)$ plus ensuring 
an overlap between the biased distributions $P_\theta(q)$ for
neighboring values $\theta_k,\theta_{k+1}$ and applying the condition 
$e^{q / \theta_k}Z(\theta_k) P_{\theta_k}(q)=e^{q / \theta_{k+1}}
Z(\theta_{k+1}) P_{\theta_{k+1}}(q)$ for those values $q$ in the overlapping
region, i.e., where 
one has good statistics for both $P_{\theta_k}(q)$ and $P_{\theta_{k+1}}(q)$.
Clearly if several values of $q$ are contained in the overlapping region,
the condition can only be approximately fulfilled simultaneously,
e.g., by minimizing the mean-square deviation. 
For details see 
Refs.~\cite{align2002,graph,work}.

\section{\label{sec:results}Results}

We have studied the Nagel-Schreckenberg model for system sizes between
$L=100$ and $L=2000$, for eighteen different car densities \change{ $\rho\in [0.02,0.8]$}.
  We used a maximum velocity $v_{\max} = 5$
and a breaking probability $p = 0.2$. All histories $\underline{Y}$, 
for each value 
of $\rho$ started with a precomputed steady-state configuration 
$\underline{y}^{(0)}$ of the corresponding density. For each history,
the Nagel-Schreckenberg dynamics was performed for $n=300$ times steps,
longer than the correlation time,
and the traffic flow $q$ of the final configuration was measured.
To obtain the distribution of these traffic flows, we performed
MC simulations in the biased ensemble Eq.~(\ref{eq:bias}) for
several values of the temperature parameter $\theta$ 
ranging from 8 different values for $L=100$ to 28 different
values for $L=2000$. The length of the MC simulations was between
$10^4$ MC steps for $L=100$  and
$10^{9}$ steps for $L=1000$ (lowest density $\rho$). Due to long equilibration times for low densities and little finite size effects already for $L = 1000$ for high and medium densities, the model was not investigated over the full support for $L = 2000$ for very low densities. 

First, we present below the results for the distributions of the traffic
flow and for
 the corresponding rate functions. We fit the tails of the distributions
to exponential functions and analyze the behavior of the fit parameters as
a function of the car density $\rho$ and relate this to the fundamental
diagram $\langle q\rangle (\rho)$. In the second section, we seek to
establish a connection of the traffic-flow distributions to other values,
i.e., ask the question what particular properties of traffic configurations
are responsible for typical or particular small or high traffic flow.

\subsection{\label{sec:distr}Distributions and Rate Functions}

\begin{figure}[!ht]
\centering
 \subfigure[]{\includegraphics[width = 0.9 \columnwidth]{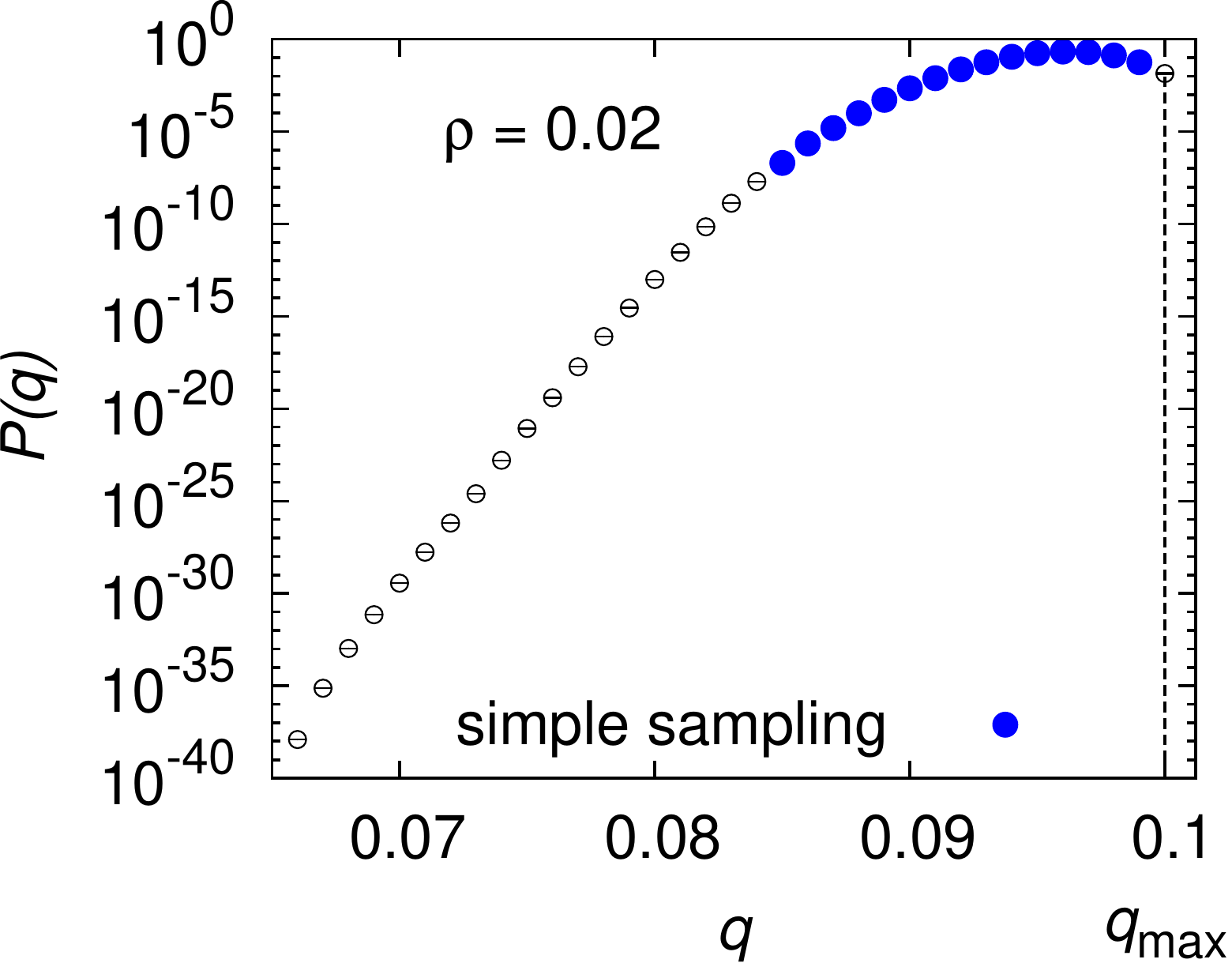} \label{fig:r002} }
 \subfigure[]{\includegraphics[width = 0.9 \columnwidth]{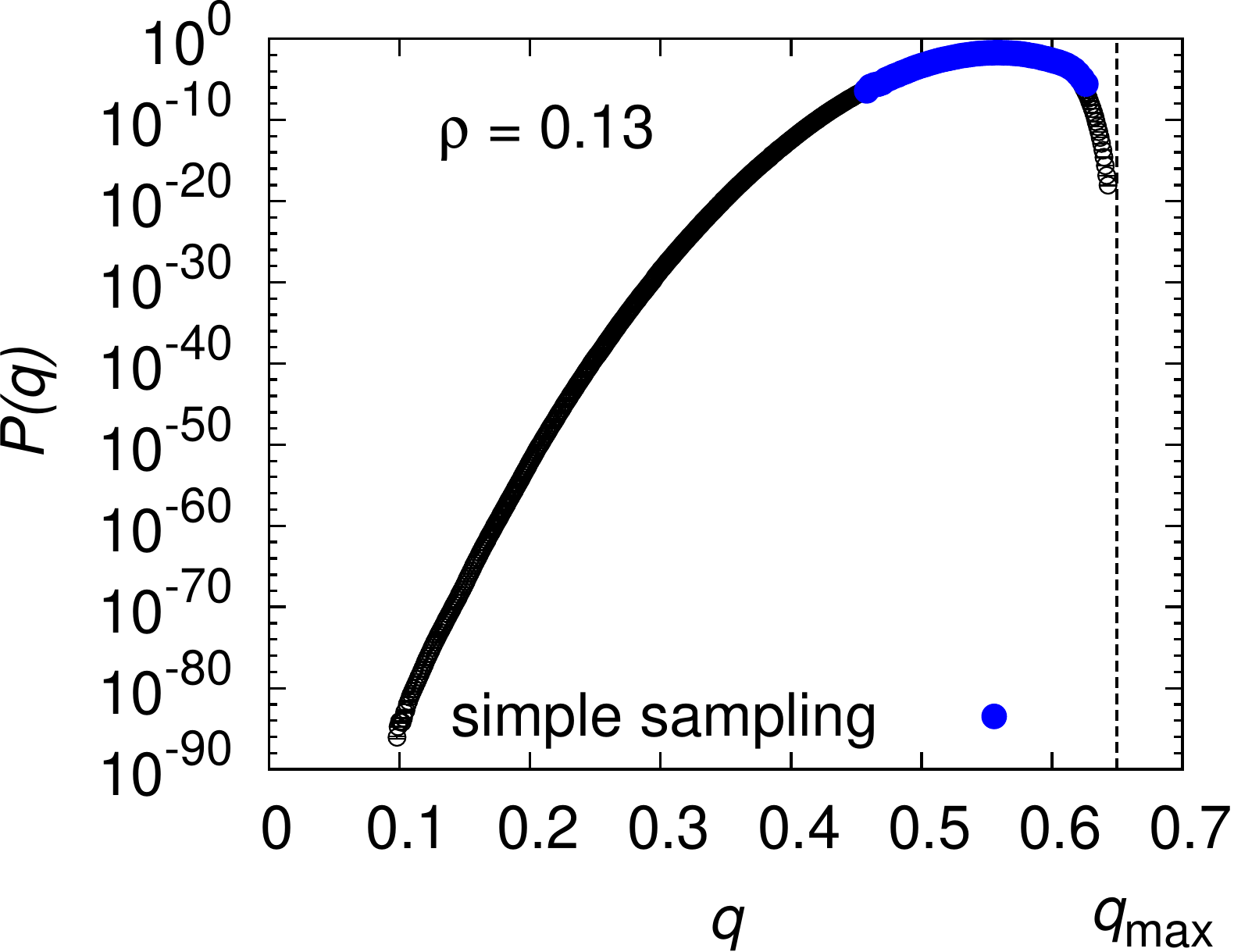} \label{fig:r013} }
 \subfigure[]{\includegraphics[width = 0.9\columnwidth]{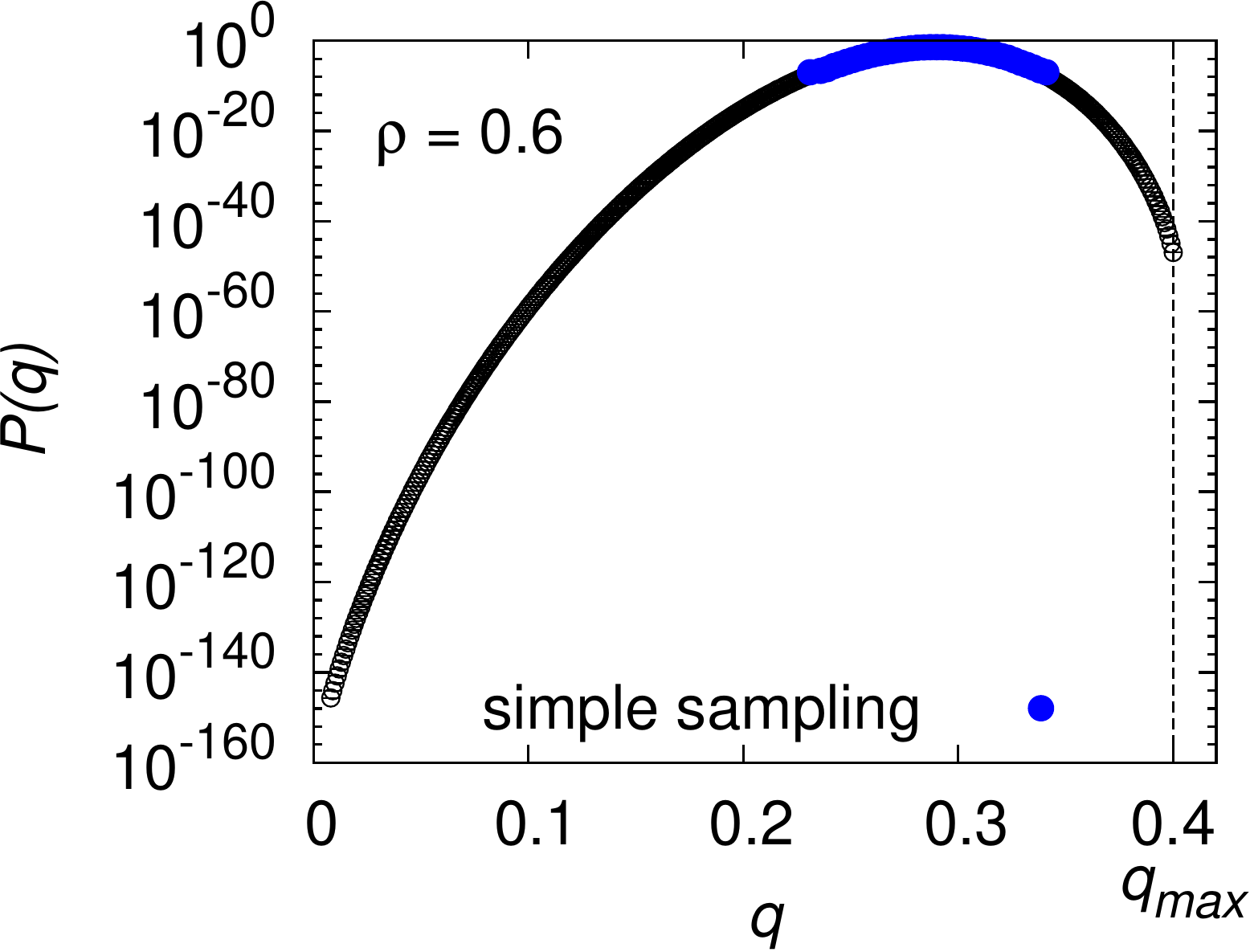} \label{fig:r06} }
\caption{(color online) Probability distributions of the traffic flow $q$ 
for $L = 1000$ for three different 
densities $\rho = 0.02 < \rho_{\max}$ (a), 
$\rho = 0.13 \approx \rho_{\max}$ (b) and $\rho = 0.6>\rho_{\max}$ 
(c). The blue larger symbols indicate the result from
simple sampling simulations. The vertical line at $q_{\max}$ indicates 
the maximum possible flow as given by Eq.~(\ref{eq:qmax})} 
\label{fig:pdf}
\end{figure}

Figure \ref{fig:pdf} shows the distributions $P(q)$ of the traffic
flow $q$ for a system of size $L = 1000$ 
for three different representative densities $\rho$. 
Figure \ref{fig:r002} shows the
behavior of $P(q)$ in the low density regime, here $\rho=0.02$,
where a free flow is dominating.
 The simple sampling results are denoted by blue slightly large symbols. 
For the right tail, using our approach we were able to sample
till the maximum possible flow.  
For small densities  it is
possible that all cars drive with maximum velocity $v_{\max}$. This
determines the maximum via velocity times density.  Note that
for quite high densities
this does not work any more and the maximum flow is determined by assuming that
all free lattice sites are ``visited'' by a car, i.e., by the
fraction of free lattice sites:
\begin{equation}
q_{\max}=\left\{
\begin{array}{rl}
v_{\max}\rho & \quad \mbox{for} \, \rho \, \mbox{small}\\
1-\rho & \quad \mbox{for} \,\rho \, \mbox{large}
\end{array}
  \right.\,.
\label{eq:qmax}
\end{equation}

For the left tail, even though not the full distribution was
obtained in this regime due to too long equilibration times of the
MC simulations, still a considerable
large part of the support could be sampled by
applying the large-deviation approach,
 down to probabilities like $10^{-38}$.
 In general, the results show, that the distribution
is dominated by a peak very close to the maximum flow, but rare events 
with smaller flows occur.

Figure
\ref{fig:r013} shows the distribution near the transition. Because $q_{\max}$
is larger, the distribution covers
a much larger range of the support compared to the low density result.
Therefore, also much lower probabilities like $10^{-80}$ can be addressed.
The distribution has a
sharp bend in the typical regime and is stronger separated
from the upper bound $q_{\max}$.  Apart from that, the left tail
already looks quite similar to the left tail of the distribution in the
low-density case. 

A typical distribution in the congested regime is
displayed in Figure \ref{fig:r06}. Also in this regime,
the distribution is asymmetric and exhibits a cut-off at the maximum
flow $q_{\max}$. Nonetheless, the typical regime is even further apart from
this value (as compared to lower densities) 
and hence the right tail covers a larger range of the
support.

In case the distribution exhibits the properties of
the \emph{large deviation principle} 
\cite{denHollander2000,touchette2009,dembo2010,touchette2011}
this roughly means that the probabilities are in the leading
order exponentially small
in the system size $L$, i.e., $P(q)\sim e^{-L\Phi(q)+ o(L)}$. Thus, the
empirical size dependent rate function \cite{graph}
\begin{equation}
\Phi (q) = -\frac{1}{L} \text{ ln} P(q) 
\end{equation}
(we omit the $L$ dependence on the left side for simplicity)  
just picks out $\Phi$ of this leading order 
and should converge for $L\to\infty$ if the large-deviation
principle holds.
In this case, a lot of relationships
are known in large-deviation theory, such that, in principle, analytical
results are easier accessible.

\begin{figure} [ht]
\centering
 \subfigure[]{\includegraphics[width = 0.3 \textwidth, angle
 =270]{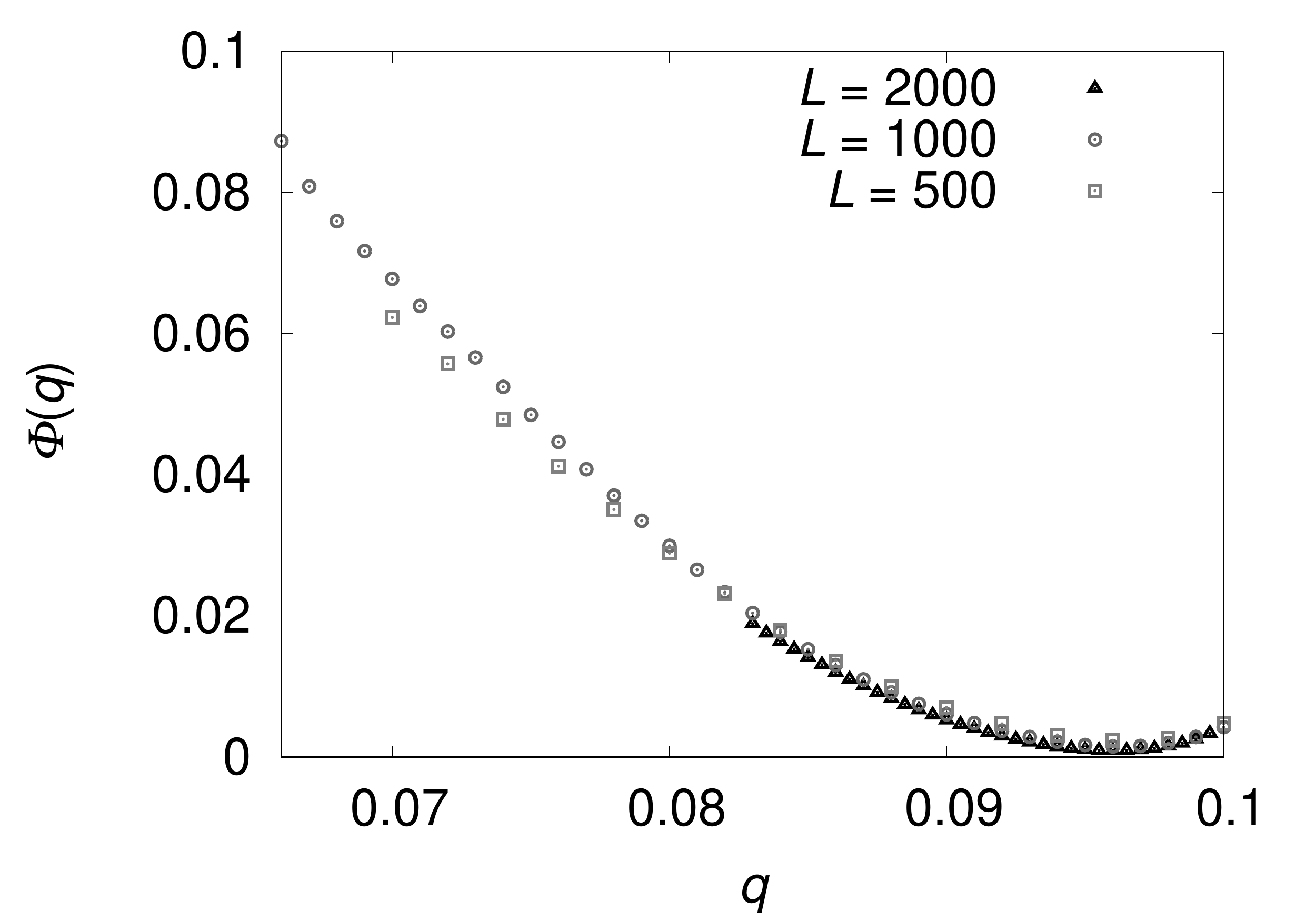} } \subfigure[]{\includegraphics[width =
 0.3 \textwidth,, angle =270]{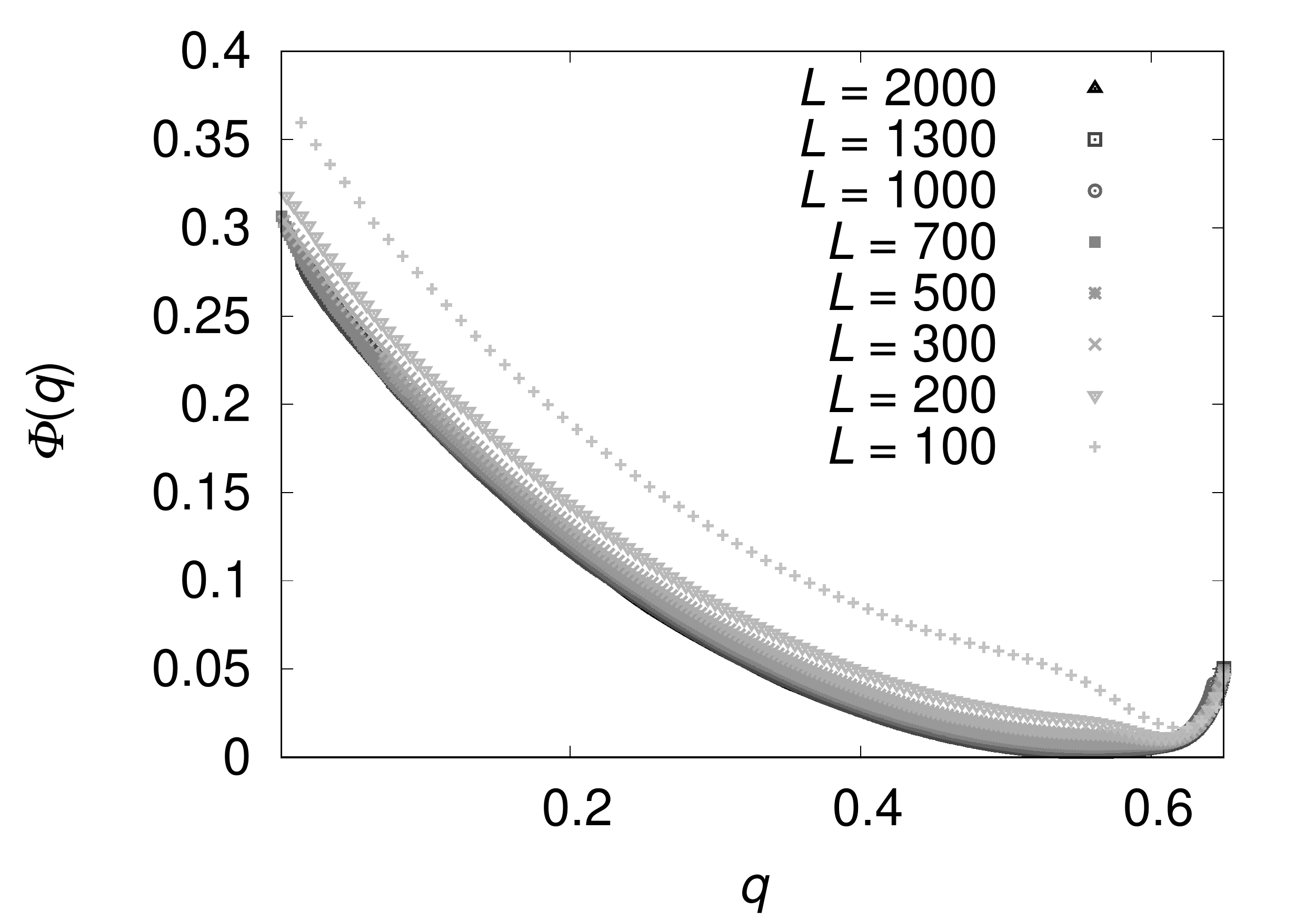} }
 \subfigure[]{\includegraphics[width = 0.3 \textwidth,, angle
 =270]{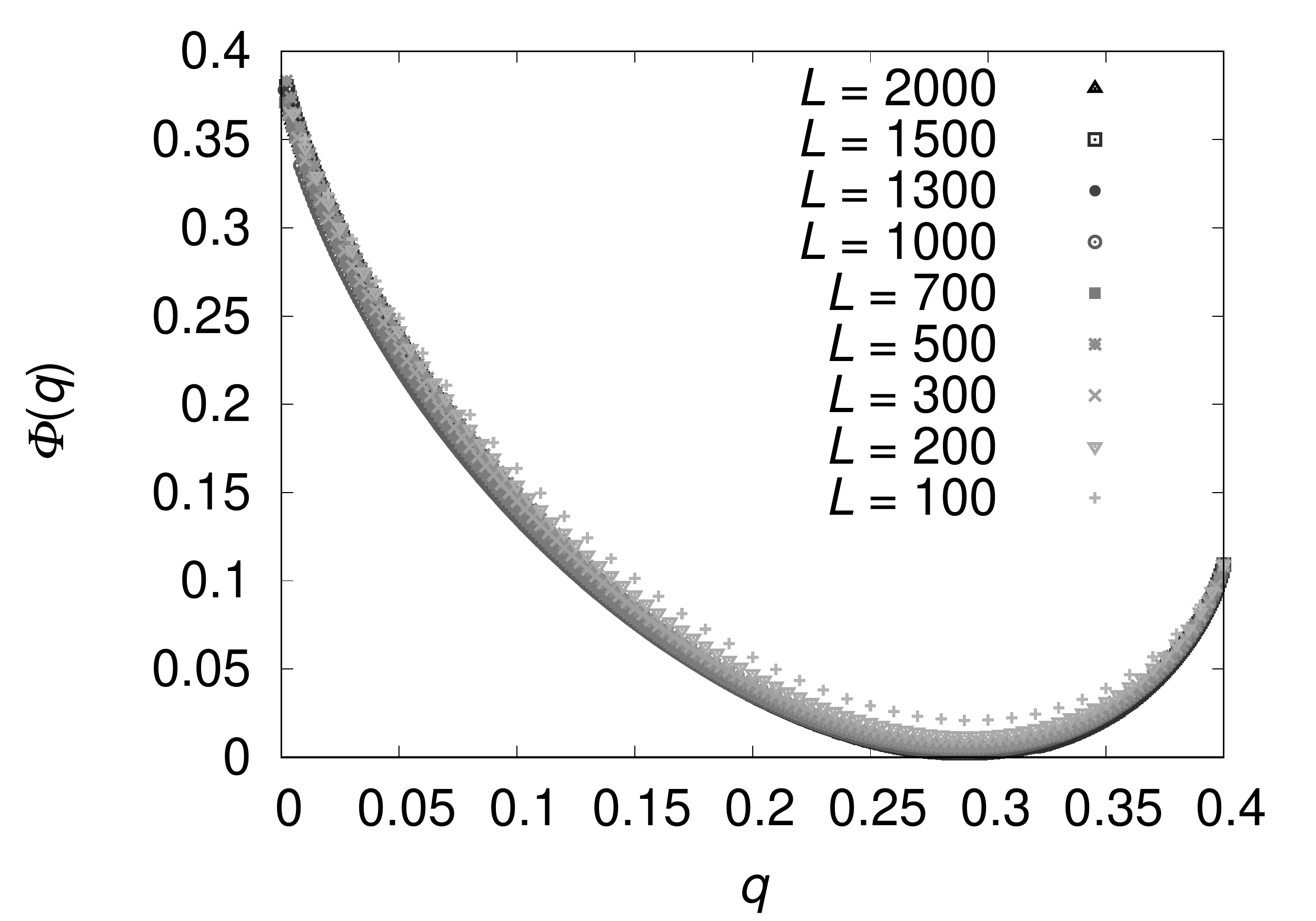} }
\caption{Empirical rate function $\Phi(q)$  for various system sizes $L$ and 
 three different densities $\rho = 0.02 < \rho_{\max}$ (a), 
$\rho = 0.13 \approx \rho_{\max}$ (b) and $\rho = 0.6>\rho_{\max}$ (c).} 
\label{fig:rate}
\end{figure}

Figure \ref{fig:rate} shows the resulting rate functions
for the same densities $\rho = 0.02$, $\rho = 0.13$ and $\rho = 0.6$,
for different system sizes $L$. For increasing value of $L$,
the empirical rate functions agree better and better \cite{graph},
indicating a convergence. 
Therefore, the large-deviation principle seems to be fulfilled for all
considered densities. The largest system sizes we can reach seem
to be large enough to observe the limiting behavior rather well.
 Note, that the finite size effects seem to have
the largest impact on the rate function for the medium density. This
is due to a shift of the maximum of the fundamental diagram towards
lower densities for larger system sizes and resembles the occurrence of 
stronger finite-size effects near phase transitions.

\begin{figure} [ht]
\centering
{\includegraphics[width = 0.98 \columnwidth]{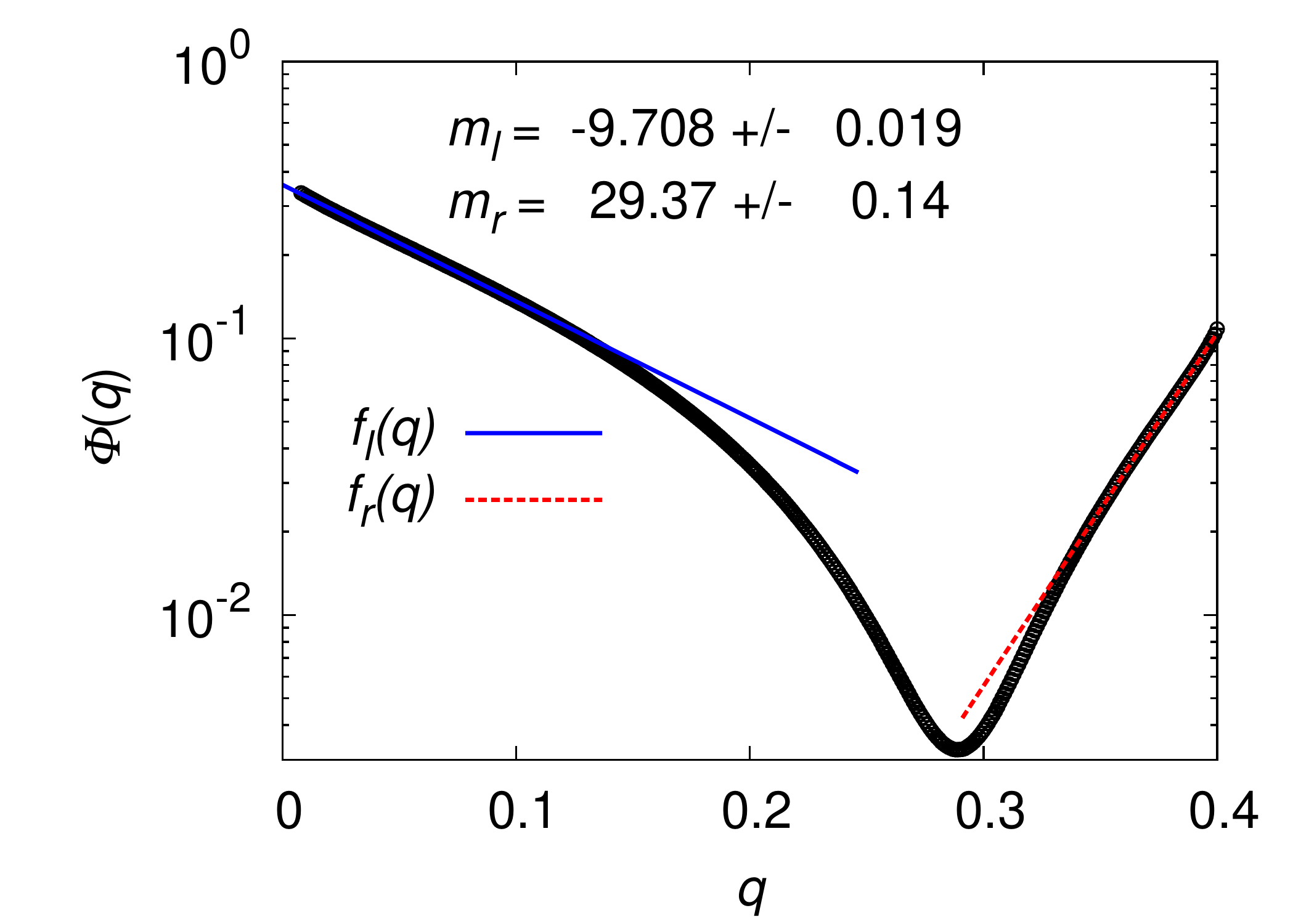}}
\caption{\label{fig:fit} 
(color online) Empirical rate function for $L = 1000$ 
and $\rho = 0.6$. An exponential functions (\ref{eq:f_r}) and (\ref{eq:f_l})
have been fitted to the tails, yielding fit-parameters $m_l$ and $m_r$
describing the shape of the
left and  the right tails, respectively.} 
\end{figure}

The shape of the rate functions hints that the tails might behave
exponentially. This is confirmed by plotting the rate function
with a logarithmically scaled $y$-axis, as shown in Figure \ref{fig:fit}.
Therefore 
we have fitted the functions
\begin{equation}
\label{eq:f_r}
f_r (q) = e^{C_r + m_r\cdot q}
\end{equation}
with fit parameters $C_r, m_r \in \mathbb{R}$ to the right tails and
\begin{equation}
\label{eq:f_l}
f_l(q) = e^{C_l + m_l\cdot q}
\end{equation}
with fit parameters $C_l, m_l \in \mathbb{R}$ to the left tails
of the rate functions data.
 The aim of this approach is to obtain fit parameters $m_r(\rho)$ and
 $m_l(\rho)$ for different densities $\rho$ and investigate thereby
 whether and how the transition from low (free flow) to high density 
(congested) regime
 affects the shape of the large-deviation tails. 
While the behavior in the typical
 regime is already well known by previous investigations of the
 fundamental diagram \cite{SchadRev}, the impact of the
 transition on the entire distribution of the traffic flow $q$ has (to
 the authors' knowledge) not been explored yet at all. Thus, to also
 investigate the atypical regime, the large-deviation tails are of
 particular interest. Figure \ref{fig:fit} also shows examples for the
 obtained fit functions.

\begin{figure} [ht]
\centering
{\includegraphics[width = 0.32 \textwidth, angle
=270]{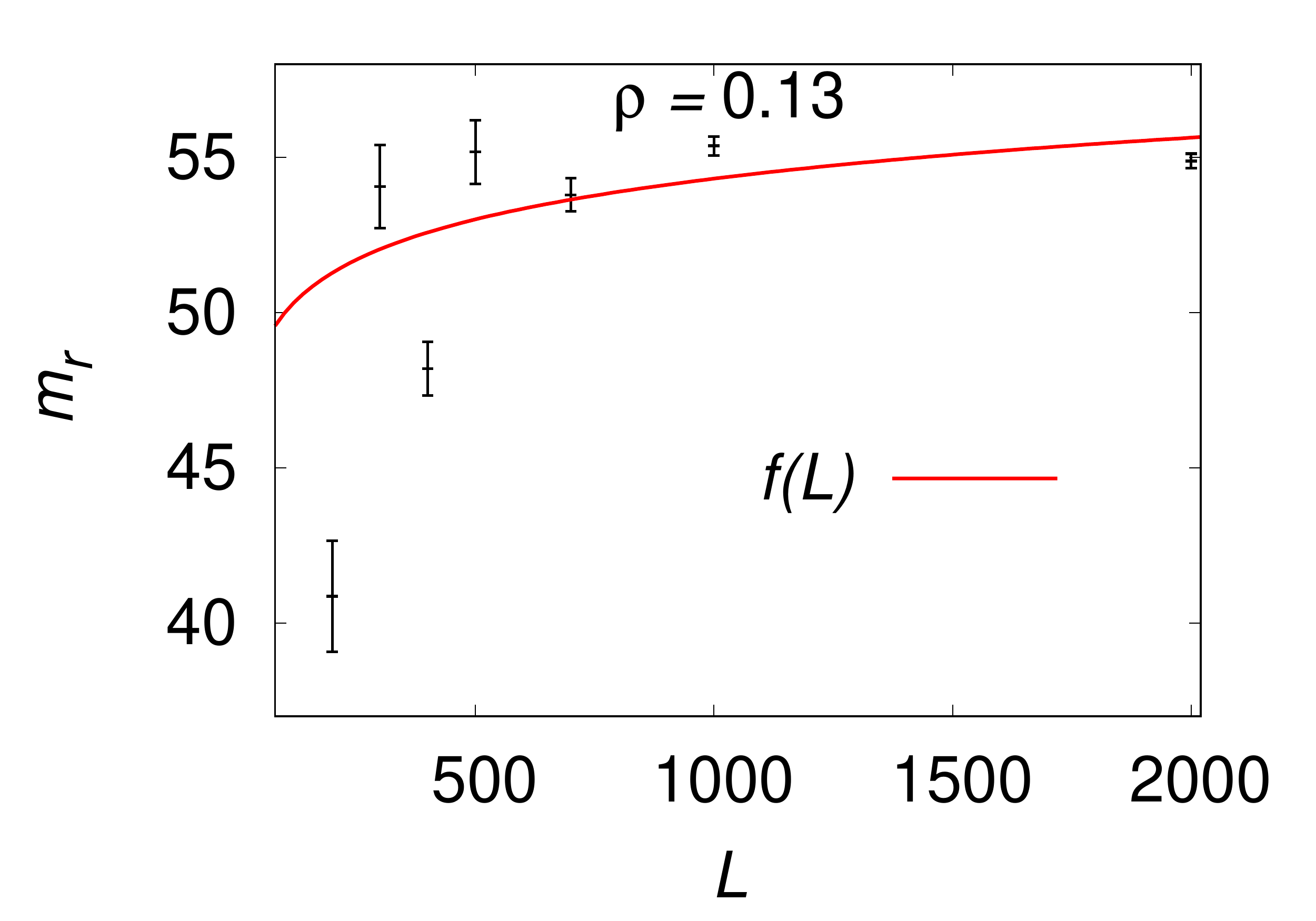} }
\caption{\label{fig:Lfit} Fit parameter $m_r$ to the right tail as a function of the system size for a density $\rho = 0.13$}
 \end{figure}

In order to investigate the
system-size dependency of the fit parameters $m_l$ and $m_r$, we 
performed for two typical  densities $\rho=0.13$ and $\rho=0.6$
the fits to the tails of the empirical rate function  for all 
available system sizes
$L$. Thus, for each size $L$, the parameters $m_l$ and $m_r$ 
were obtained.  As an example, the parameter $m_r$ is shown for one density
in Figure \ref{fig:Lfit} as a function of $L$. Although we observe quite
a bit of scatter for small system sizes $L$ (due to the fact that
we display not directly measured values but fit parameters), 
the result does not change
much for large system sizes, indicating a convergence for $L\to\infty$.
To investigate this quantitatively, the function
\begin{equation}
\label{eq:fit}
f(L) = a+c\cdot L^b
\end{equation} 
is used as fit function for the data displayed in Figure
\ref{fig:Lfit}, resulting in the displayed curve.
The fit is not very good, which is anyway unavoidable
due to  the scatter of the data and the small error bars which are 
purely statistical.
Nevertheless, for the shown example as well as for $m_l$, both for
the two considered values of $\rho$, 
the values we have found for the largest system do not differ considerably
from the extrapolated values (in terms of the general order of magnitude
and the obtained error bars; although the actual values should not be taken too
seriously). Thus, we just take the values of $m_r$ and $m_l$ 
obtained from the fits to the largest system sizes as good representatives
of the exponential tail behavior of the rate functions for all
eighteen different values of $\rho$ we have studied.

\begin{figure} [ht]
\centering
{\includegraphics[width = 0.98\columnwidth, angle =
270]{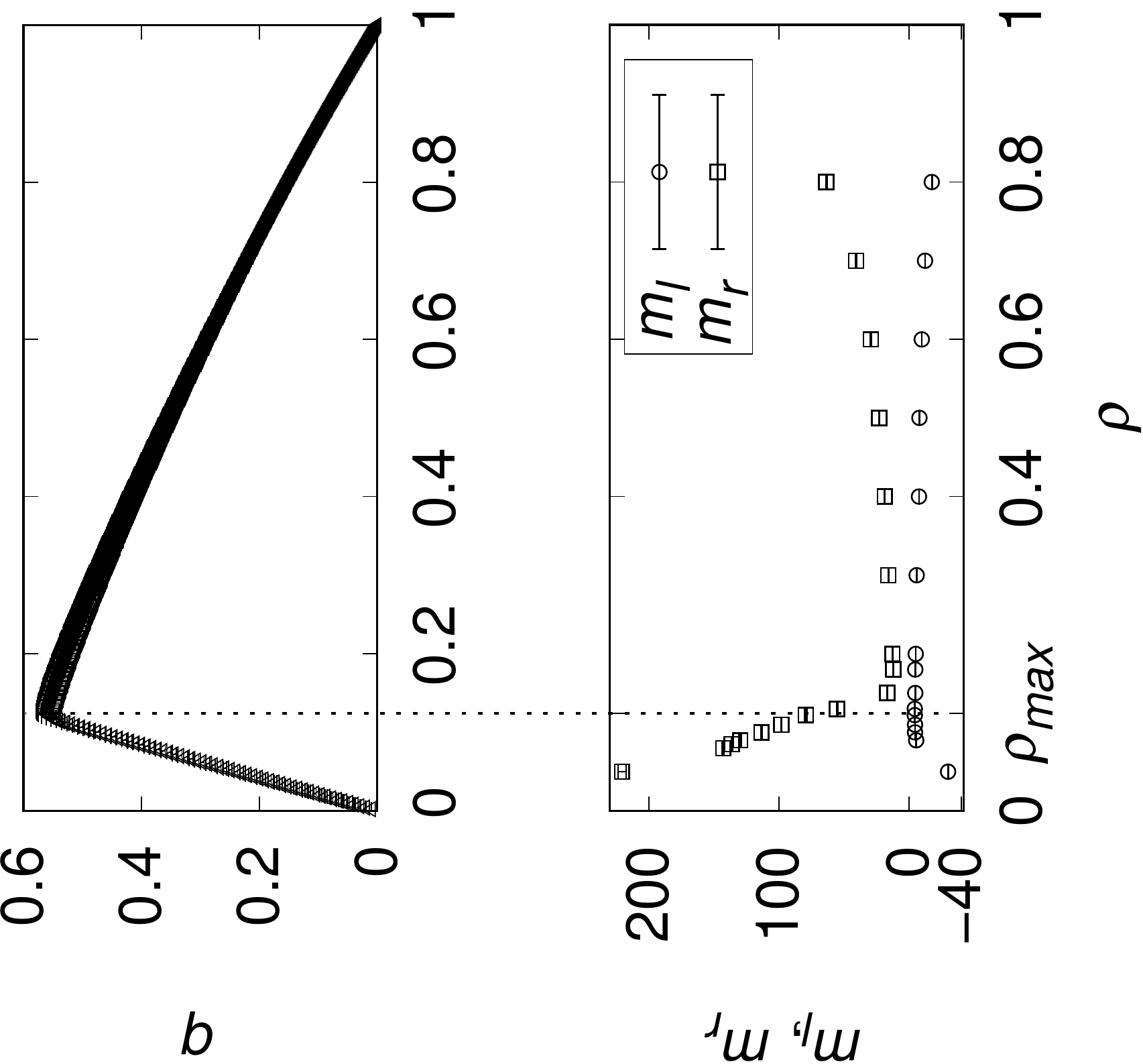} }
\caption{\label{fig:fund_para} Fundamental diagram (upper picture). The density of maximum flow $\rho_{\max}$ is marked by the dashed line. Fit parameters $m_l$ and $m_r$ as a function of the density of cars $\rho$ (lower picture). Both parameters exhibit an extreme value near the density of maximum flow. }
 \end{figure}

 Figure \ref{fig:fund_para} shows the fundamental diagram (upper
 picture) and the fit parameters $m_l$ and $m_r$ for both tails (lower picture)
as a  function of the density of cars $\rho$. The maximum
 of the fundamental diagram is marked by a vertical dashed line. Both
 fit parameters exhibit an extreme value near the density $\rho_{\max}$ at
 which this maximum occurs.
 
A closer look reveals that the minimum of the fit parameter $m_r$ 
appears  at a slightly higher density than the
density $\rho_{\max}$ that maximizes the flow in the fundamental diagram. This
reminds us of previous results \cite{Krug} obtained for the dissolution
time $\tau_J$ of an initial mega jam, which shows a sharp increase near
the critical density $\rho_c = \frac{1}{v_{\max} + 1}$ at which the
system exhibits a phase transition in the deterministic limit $p = 0$.

\begin{figure} [ht]
\centering
{\includegraphics[width = 0.8 \columnwidth]{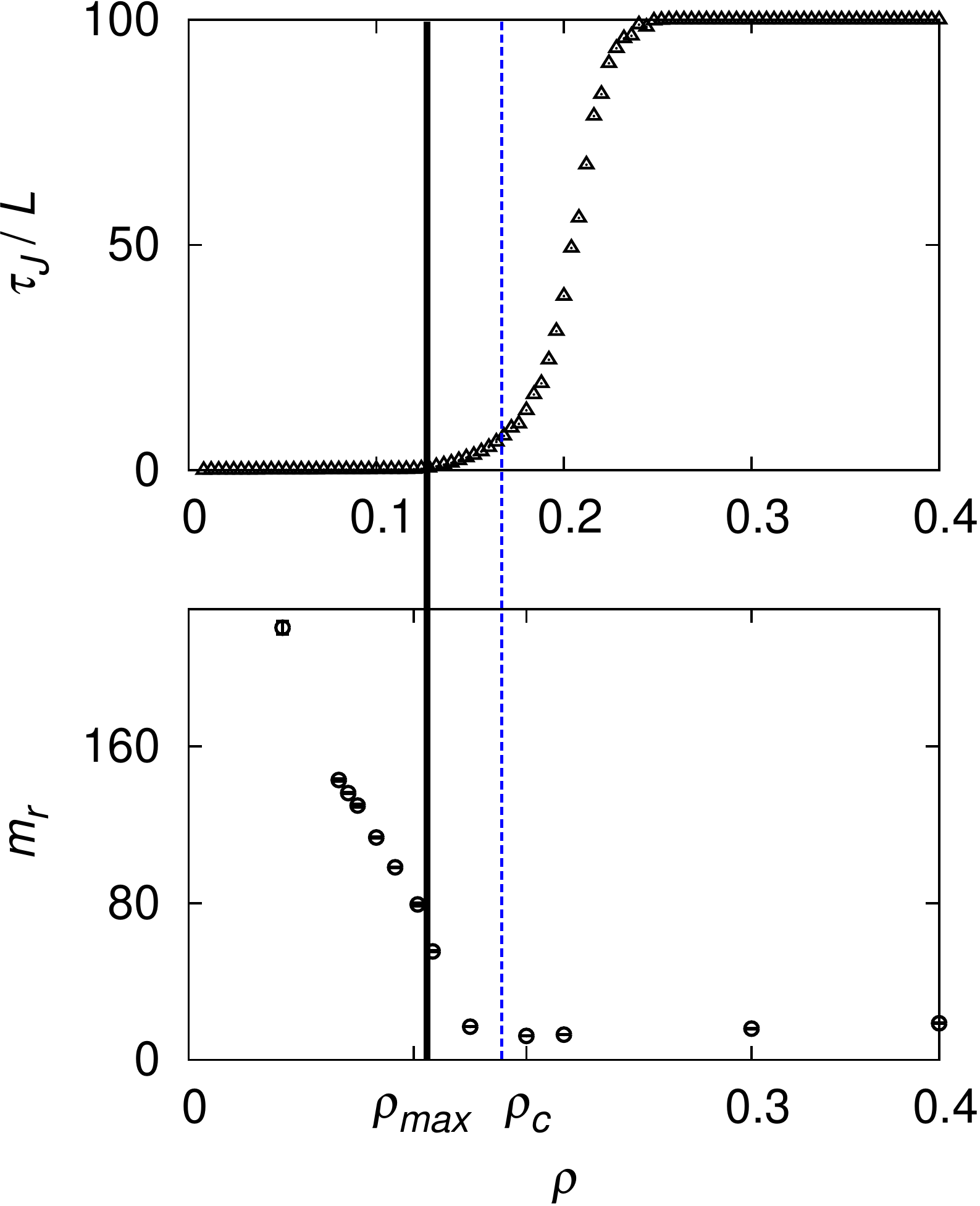} }
\caption{ \label{fig:Jam_diss_para}
The upper picture shows the dissolution time $\tau_J$ of an 
initial mega jam. The density $\rho_{\max}$ that marks the 
maximum of the fundamental diagram and the critical density 
$\rho_c$ of the deterministic limit $p = 0$ are marked by 
dashed lines. The lower picture shows the fit parameter $m_r$ 
as a function of the density of cars $\rho$. }
 \end{figure}

This motivated us, to compare our results also with this 
jam dissolution time $\tau_J$ (which we obtained
from separate standard simulations), which is shown in the 
upper picture of Figure \ref{fig:Jam_diss_para}
 as a function of the density
 of cars $\rho$. The lower picture shows the fit parameter $m_r$ for
 the same range of densities $\rho$. In both cases, the same system of
 size $L = 1000$ was used. One can see that the position of the
 minimum of $m_r$ lies in the range where $\tau_J$ increases
 strongly and closer to the 
the critical density $\rho_c = \frac{1}{v_{\max}+1}$ of
the deterministic limit. A reason for this behavior is not obvious to us
in the moment.

\begin{figure} [ht]
\centering
{\includegraphics[width = 0.8 \columnwidth]{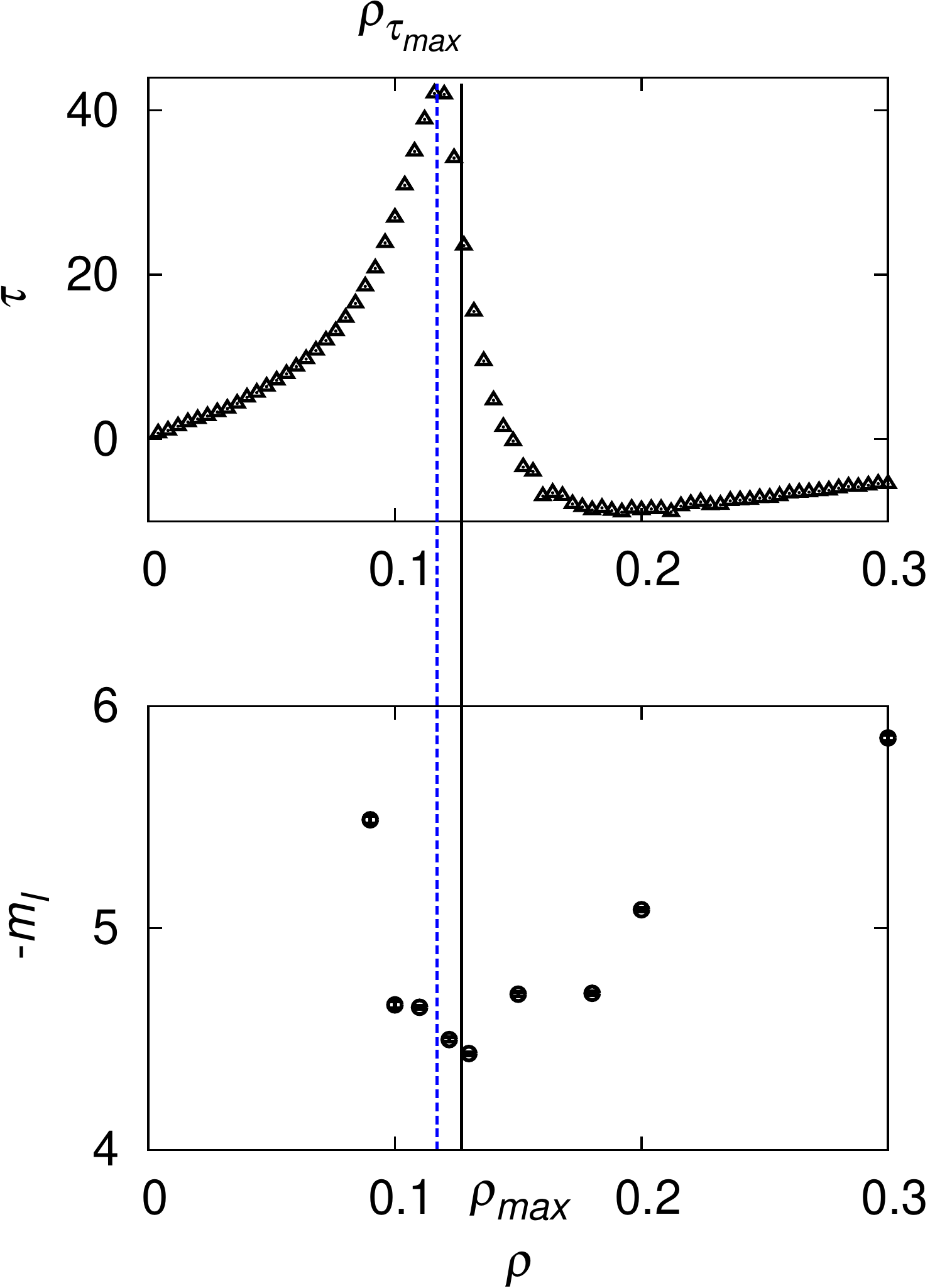} }
\caption{\label{fig:Relax_Para} 
The relaxation time $\tau$ (see Ref.~\cite{relax}; upper figure) and the
fit parameter $m_l$ for the left tail of the distribution
function as a function of the car density $\rho$.
The density $\rho_{\max}$ that marks the maximum of the
fundamental diagram and the density $\rho_{\tau{\max}}$ at which $\tau$
exhibits its maximum are marked by dashed lines.}  
\end{figure}

Next, we have a closer look at the parameter $m_l$, as displayed
in Figure \ref{fig:Relax_Para}. 
This parameter also exhibits an extreme value, which is, as
mentioned above, 
at another density than the minimum of the fit parameter $m_r$.
 Since the relaxation time $\tau$ (see \cite{relax}) is known to
 exhibit a maximum at a density below $\rho_{\max}$, it is displayed in
 the upper picture of Figure \ref{fig:Relax_Para}.
We observe (from additional simple sampling simulations) 
a maximum at some value $\rho_{\tau_{\max}}$.
Since the slope of the curve is quite low in this region
 and the scattering of the data points in comparison is rather high,
 one cannot exactly determine the peak's position precisely.
 But it appears to be located slightly below
the value of $\rho_{\max}$ in agreement with the previous results \cite{relax}.
Nevertheless,
with the current precision of our data we cannot decide whether
the maximum of $m_l$  coincides rather
 with the maximum of the relaxation time $\rho_{\tau_{\max}}$ 
or with the position of maximum
 of the fundamental diagram $\rho_{\max}$, both appears to be possible.
Nevertheless, the general results holds, that the tails of the
distribution change considerably closely to the density where the 
change from the free-flow to the congested regime occurs.

\subsection{\label{sec:corr}Correlations}

Since we have access even to the extremest traffic situations which
are possible under each given circumstances, it is worth asking
whether there are other characteristics properties of these rare situations,
in addition to extreme small or large values of the flow $q$. Thus,
during our simulations we also measured and stored other quantities of 
interest, as stated below.
This allows us to  correlate them within our analysis
with the observed traffic flow. Since our approach allows us to
access the tails of the distributions, we are therefore able to investigate
the correlations also over a very large range, much beyond the typical
correlations. Understanding these
correlations also for the extreme cases, might help to foster (or avoid)
(un-)desired traffic situations in real cases. We concentrate the results
we show on the region in the phase disgram near $\rho_{\max}$. 
For very low densities and very high densities, the
correlations look typically very simple (or not much different from the
presented), 
thus they are  not discussed further here.

\begin{figure} [ht]
\centering
{\includegraphics[width = 0.3 \textwidth, angle
=270]{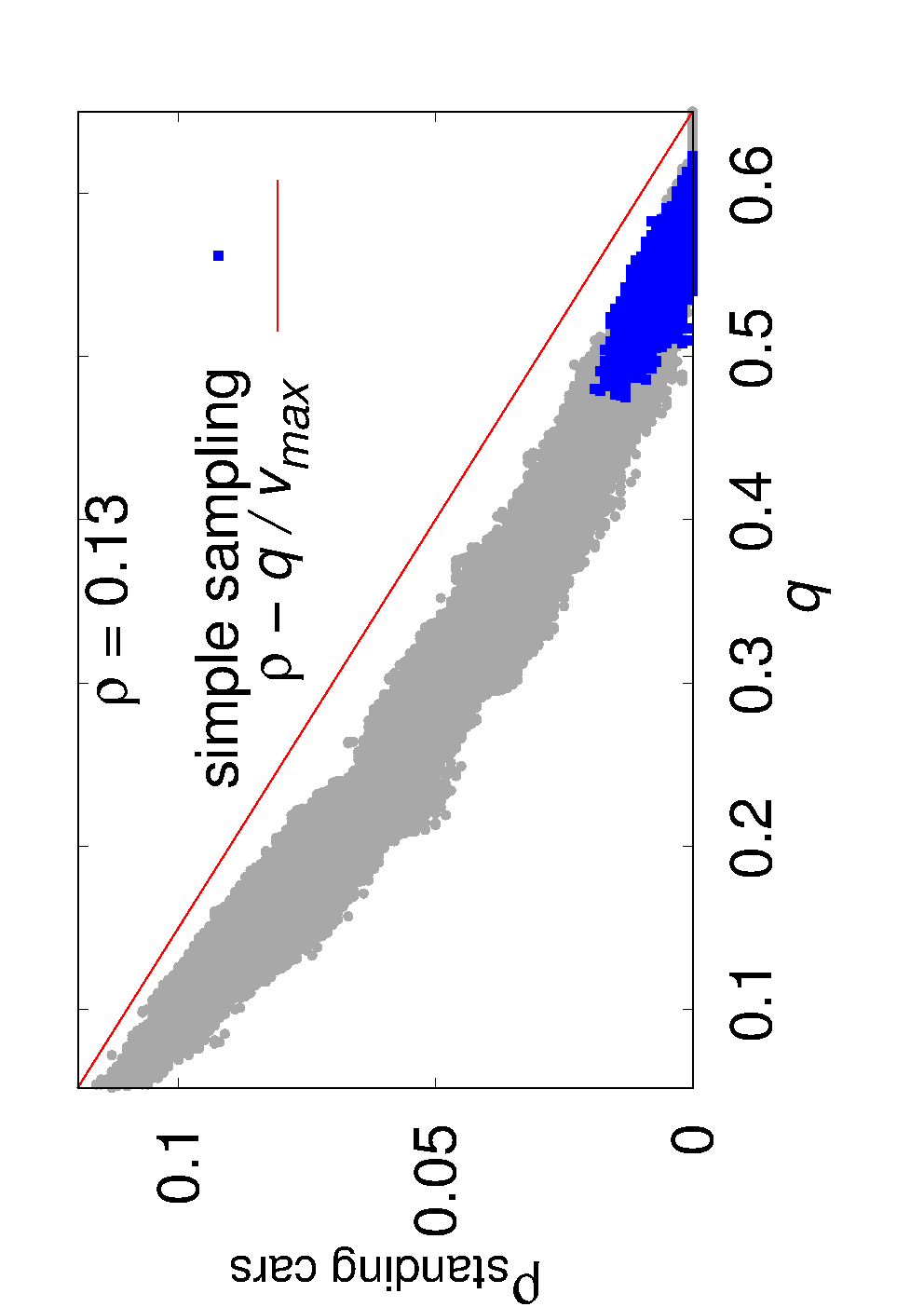} }
\caption{\label{fig:Korr_r013_stand}
(color online) Scatter plot for correlation between the traffic flow 
$q$ and the density $\rho_{\text{standing cars}}$ of cars with velocity zero for a 
system of size $L = 1000$ with $\rho = 0.13$. The dark blue points indicate the data
which can be found by simple sampling. The line indicates the maximum possible
density $\rho_{\text{standing cars}}$ as function of $q$. 
}
 \end{figure}

Figure \ref{fig:Korr_r013_stand} displays the correlation between the
traffic flow $q$ and the density $\rho_{\text{standing cars}}$ of
standing cars by a scatter plot.
The results obtained by using the simple sampling approach are marked
in blue. One can clearly see that the large-deviation approach allows
us to investigate the correlations over a much wider range of the
support.
One can formulate a simple bound for these correlations: 
At a given density $\rho_{\text{standing cars}}$ of
standing cars, the largest possible value for the traffic flow $q$
would be measured when all non-standing cars had maximum velocity
$v_{\max}$ (which often might still not be possible), hence
\begin{equation}
\label{eq:q_rho}
\rho_{\text{standing cars}} = \rho - \frac{q}{v_{\max}}
\end{equation}
is an upper bound for the density of standing cars at a given value of
$q$. 
In Figure \ref{fig:Korr_r013_stand}, 
a very strong correlation between $\rho_{\text{standing cars}}$ and $q$
is visible, with only a rather small scattering band. This means that 
the observed value of $q$ can be explained to a large extend 
by the fraction of standing cars, even in the rare-event region. Note that
the data points approach the limit Eq.~(\ref{eq:q_rho}) for $q \rightarrow 0$. The
reason for this behavior might be that a high number of standing
cars means more space for the still unjammed cars, which therefore can
have higher velocities. However, this would only be the case when the
jammed cars were in the same local area, hence if there were a few
large jams. This is indeed the case, as will be pointed out below.

\begin{figure} [ht]
\centering
{\includegraphics[width = 0.3 \textwidth, angle
=270]{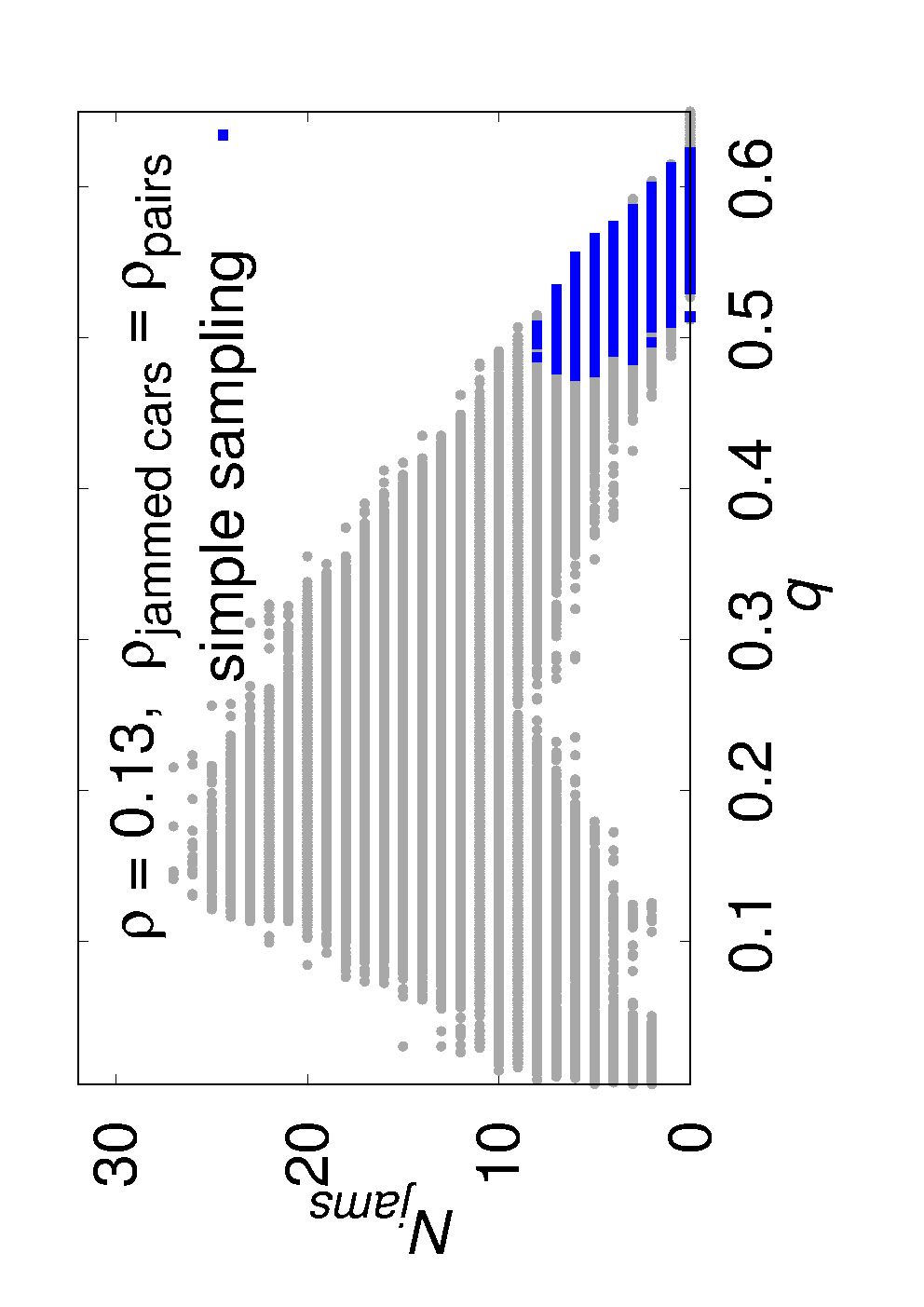} }
\caption{\label{fig:Korr_r013_N1} (color online) 
Scatter plot showing the correlation between the traffic flow $q$ and 
the number $N_{\rm jams}$ of jams for a system of size $L = 1000$ and 
$\rho = 0.13$. 
The dark blue points indicate the data which can be found by simple sampling.}
 \end{figure}

There are different ways of defining if a car is jammed, see
 \cite{SchadRev}. Hereinafter, a car is defined jammed when it had to
 brake to zero due to another vehicle ahead. Therefore, the density of
 jammed cars equals the density $\rho_\text{pairs}$ of next neighbor
 pairs.  A jam is therefore given by a cluster of standing cars with no
gaps. Correspondingly the number $N_{\rm jams}$ of jams is the number of these
clusters.
Figure \ref{fig:Korr_r013_N1} shows the relation between 
 $N_{\rm jams}$ and the traffic flow $q$. Very high
 values of the flow $q$ can only occur if the number of jams is small or even zero.
 For intermediate values of the  traffic flow a broad range of the number of jams
is possible, but typically more jams occur, with a maximum near $q=0.15$. For
 $q\rightarrow 0$, only a small number of jams can be observed. Since
 the density of jammed cars $\rho_{pairs}$ increases in this case (not
 displayed), this means that this (for this value of $\rho$) very 
extreme traffic situation is dominated by few but large jams.

\begin{figure} [h]
\centering
{\includegraphics[width = 0.4 \textwidth, angle
=270]{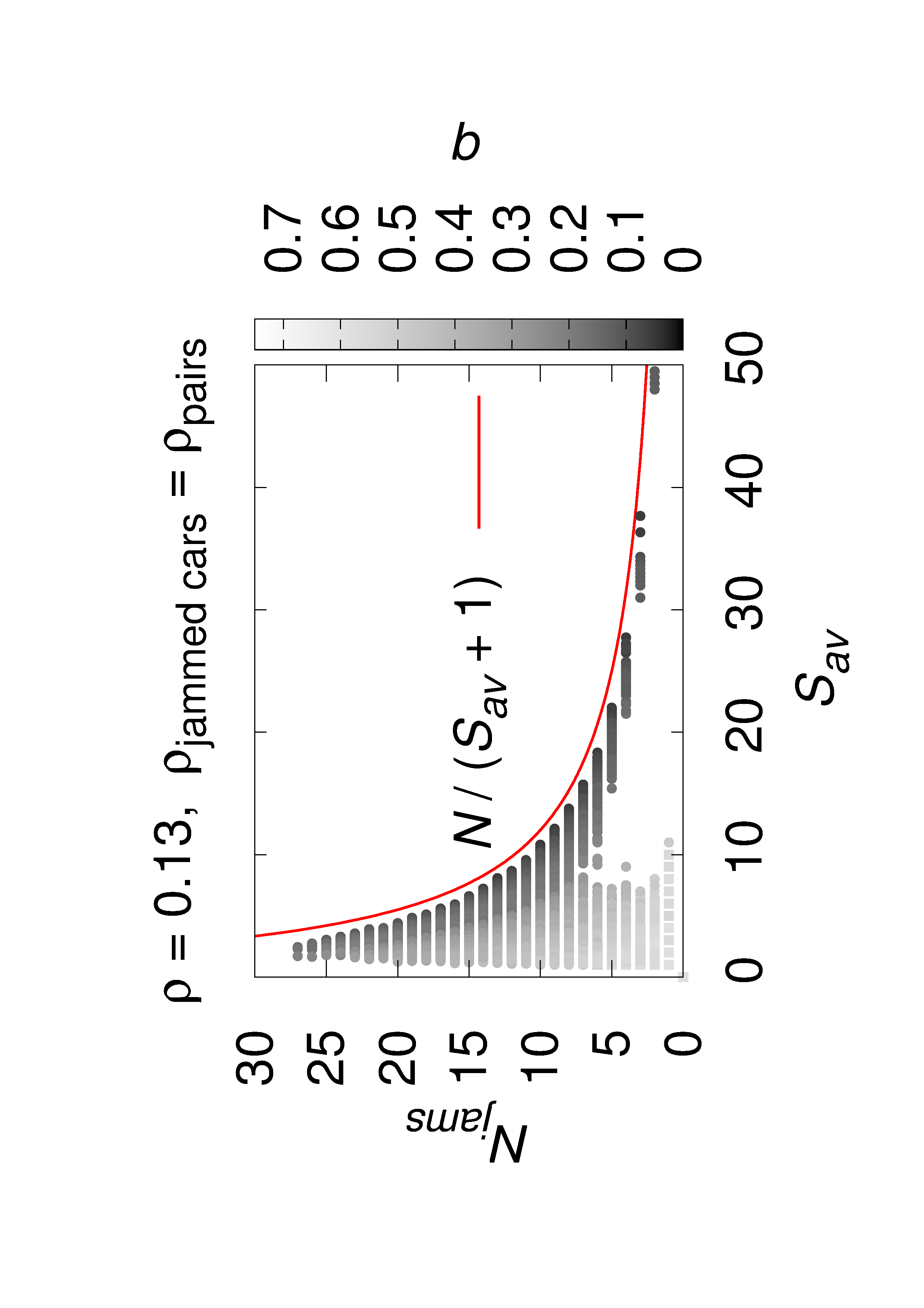} }
\caption{\label{fig:rho13_SN} 
(color online) Scatter plot showing the correlation between the number 
$N_{\rm jams}$ of 
jams and the average jam size $S_{\rm av}$ of close cars for a system of size 
$L = 1000$ and $\rho = 0.13$. The line displays the simple bound from
Eq.~(\ref{eq:limit.N.jams}).}
 \end{figure}

\begin{figure} [h]
\centering
{\includegraphics[width = 0.4 \textwidth]{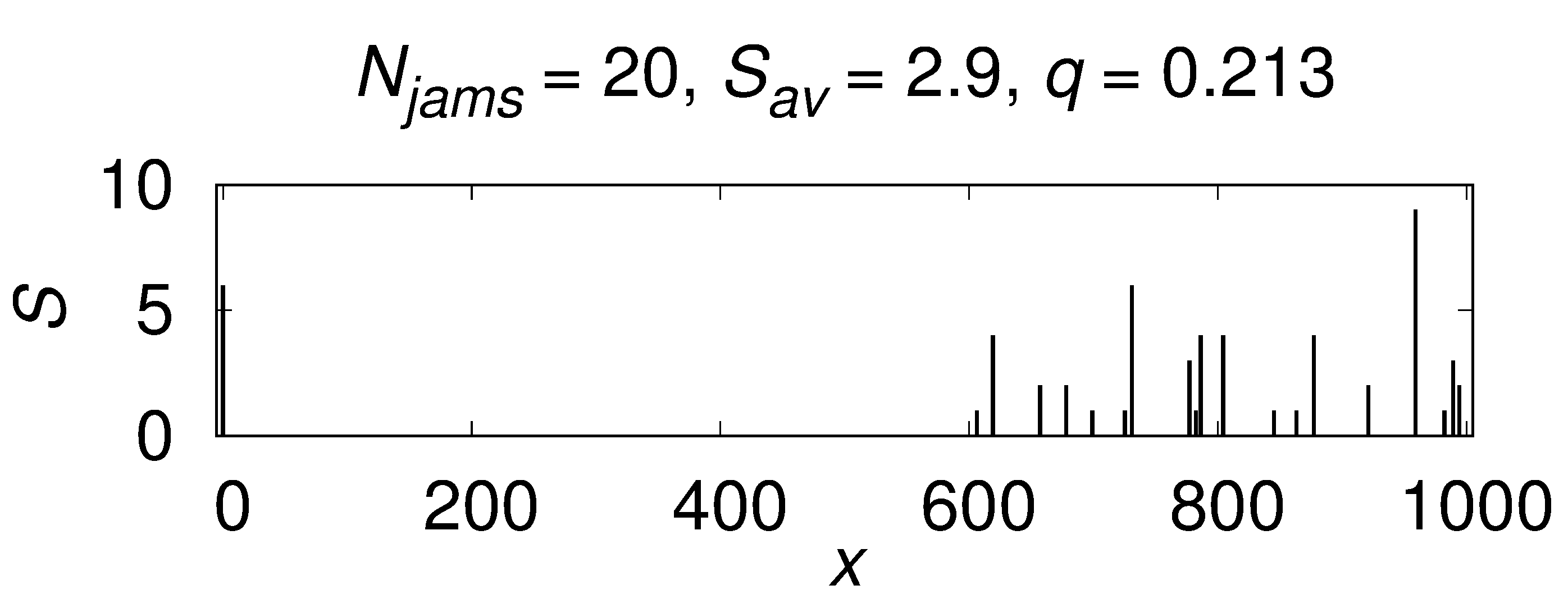} }
\caption{\label{fig:Conf_med}
Example configuration for a medium traffic flow value 
$q$ for a system of size $L = 1000$ and $\rho = 0.13$. All jams starting at position
$x$ and having size $S$ are indicated
by vertical lines at positions $x$ and of height $S$.}
 \end{figure}

\begin{figure} [h]
\centering
{\includegraphics[width = 0.4 \textwidth]{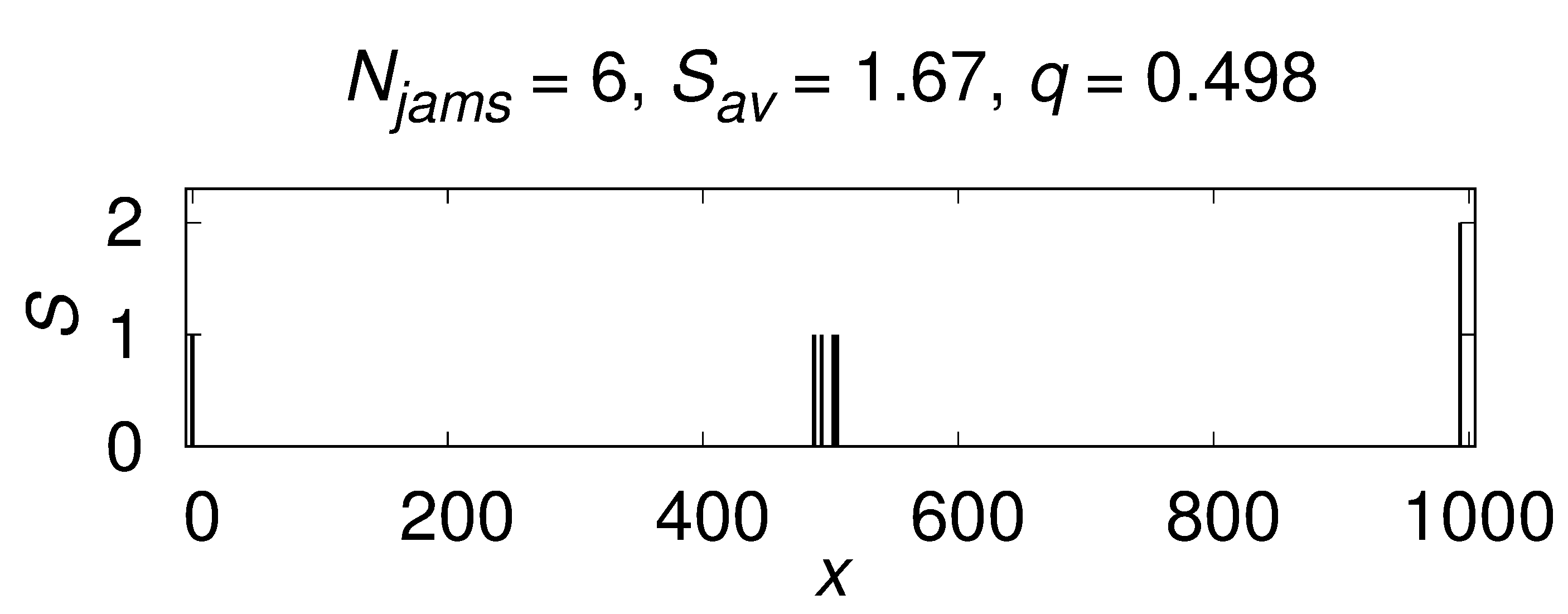} }
\caption{\label{fig:Conf_high} 
Example configuration for a high traffic flow value 
$q$ for a system of size $L = 1000$ and $\rho = 0.13$. All jams starting at position
$x$ and having size $S$ are indicated
by vertical lines at positions $x$ and of height $S$.}
 \end{figure}

\begin{figure} [h]
\centering
{\includegraphics[width = 0.4 \textwidth]{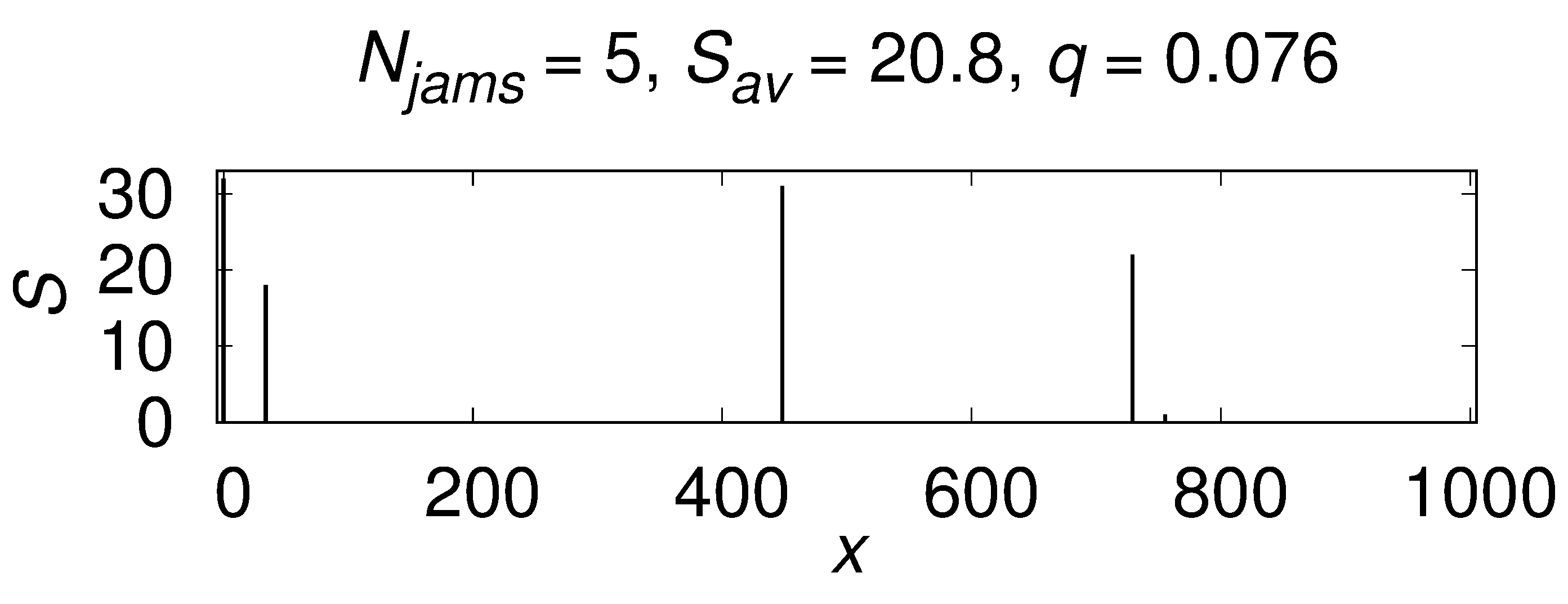} }
\caption{\label{fig:Conf_low} 
Example configuration for a low traffic flow value 
$q$ for a system of size $L = 1000$ and $\rho = 0.13$. All jams starting at position
$x$ and having size $S$ are indicated
by vertical lines at positions $x$ and of height $S$.}
 \end{figure}

To analyze the behavior observed in the preceding scatter plots, we show in 
Figure \ref{fig:rho13_SN} the correlation between the number
$N_{\rm jams}$ of jams and the average jam size $S_{\rm av}$. The average traffic flow
$q$ for any pair $(S_{\rm av},N_{\rm jams})$
is coded by a gray scale. Again, we can derive a simple bound: 
If $S_{\rm av}$ is the average number of next
neighbor pairs, $S_{\rm av} + 1$ is the average amount of cars contained in jams 
(clusters) originating from these
pairs. Since the maximum amount of jammed cars is the total number of
cars $N$, the number $N_{\rm jams}$ of jams is limited by
\begin{equation}
N_{\rm jams} = N / (S_{\rm av}+1).
\label{eq:limit.N.jams}
\end{equation}
This boundary is marked by a red line in Figure \ref{fig:rho13_SN}.
Accordingly, data points close to this line correspond to low values
of the traffic flow $q$. The observed reverse "U"-shape of the
correlation plot between the number $N_ {\rm jams}$ of jammed cars and the
traffic flow $q$ from Figure \ref{fig:Korr_r013_N1} can also be
observed here. Moreover, when being presented this way, one can
directly see that 
a large number of jams corresponds to
small jam sizes and intermediate values of $q$. An example traffic configuration
is visualized in Figure \ref{fig:Conf_med}. Furthermore 
a low number of jams can either correspond to a low
average jam size
 and a high traffic flow $q$, see example configuration in Figure \ref{fig:Conf_high},  
or to a large average jam size and a low traffic flow $q$, see 
Figure \ref{fig:Conf_low}. 

\section{\label{sec:discussion}Discussion and Conclusion}

We have studied the distributions of traffic flow $q$ for the
Nagel-Schreckenberg model, i.e., for a non-equilibrium model, for
various values of the car density $\rho$.  By applying a sophisticated
large-deviations approach, we are able to obtain $P(q)$ over a large
range of the support, down to probabilities as small as
$10^{-140}$. First, this is satisfying from a fundamental research
point of view, because only knowing the full (or almost full)
distribution contains the full information for a stochastic system.
Second, the approach allows to investigate the reasons of the occurrences of large
deviations by correlating the main quantity of interest, here $q$,
to other measurable values, like the number or sizes of jams.
 Third, also in real traffic systems
seemingly rare events may occur. Give the fact that there are millions
of kilometers of roads worldwide \cite{cia_factbook} and that traffic
situations change from minute to minute several times, a rough
estimates shows that within a year even events exhibiting a
probability like $10^{-13}$ actually occur on a global scale. Events
exhibiting such probabilities are also difficult to simulate using a
standard approach. Thus, for practical purposes also medium-small
probabilities which are accessible using the large-deviation approach
might be of practical interest.

In the present work, we have analyzed the distribution of the flow,
starting from a steady-state configuration after a rather long
evolution of $n=300$ time steps. We have shown that on the level of
the distribution of the flow, the initial configuration of cars and
speeds is mostly forgotten. Thus we basically analyzed the
large-deviation properties of the steady state, at least for
low-enough densities.

Note that it could be indeed also of interest
to study deliberately the distributions of flow after a smaller number $n$
of steps, to investigate the statistical properties of the traffic evolution
over small time intervals, and correspondingly the dependence on the
initial configuration which should be more pronounced.
This could be an interesting aspect for further studies.

The fundamental question 
of the present work is whether there is a correspondence
between the shape and large-deviation properties 
of the distribution of the flow (order) parameter
and the phase or state the system is in. Such correspondences have
been observed previously for phase transitions in 
equilibrium models, like the percolation transition of random graphs
\cite{rare-graphs2004,graph,distr_qcore2017,diameter2018,biconnected2019}. 
For the 
Nagel-Schreckenberg
model, we find exponential left and right tails in $P(q)$ for all cases,
but the slope of the tails changes significantly. The distribution
exhibits the smallest slopes
 near the change from free flow to the congested phase.
Thus, also here, for a non-equilibrium model, we see such a strong
relationship between phase and shape and tail properties of the distribution
of the ``order parameter''.

Furthermore, our correlation analysis reveals that in the onset of the 
congested  region near the
``critical'' density $\rho_{\max}$, non unexpectedly, 
 the correlation between the flow and number of standing cars is 
very straightforward (this holds also for other values of $\rho$).
Nevertheless, with respect to the number of traffic jams, the situation is more
complex. Although a large number of jams, which occurs for this density with a small
probability, corresponds to an already significantly reduced flow, a smaller number
of jams can occur for large flow (small jams), which is typical for this
density, and for small flow (large jams), which is an extreme event here.

For the present study we started always with a typical
steady-state configuration as initial configuration of any history.
Here, e.g., traffic jams are very unlikely for very small densities $\rho$
(and become more and more likely when increasing $\rho$).
Thus, it could be interesting to use other more atypical, but still
realistic, initial
configuration, which may be seen as precursors of traffic jams or other
events. Then one could again analyze the distributions $P(q)$
which will probably lead to increased traffic jam probability.
This could lead to a more precise prediction of unlikely traffic
situations, conditioned to different initial configurations.

Also, the traffic flow is not the only quantity of interest, where
one can apply the large-devaition approach. Also for other
quantities like the number or the duration of traffic jams it could
be of interest to obtain the probability distribution over a large
or full range of the support. Clearly, these distributions can not
be read off from the present results, because independent MC simulations
with biases driven by the quantities of interest have to be used.

Finally, it is obvious that we have applied the large-deviation
approach to a very simple traffic model, e.g., it exhibits just one
lane, no in- or outflow, a homogeneous type of cars and drivers.
We have chosen such a very simple model to provide a proof of principle, to
show that the large-deviation approach allows for insights beyond
those accessible by standard simple sampling.
Hence, for further studies it would be certainly interesting to apply
our approach also to more complex, i.e., more realistic traffic models.

\begin{acknowledgments} 
The simulations were performed on the HERO and CARL clusters of the University
of Oldenburg jointly funded by the DFG (INST 184/108-1 FUGG  
and INST 184/157-1 FUGG) and the 
ministry of Science and Culture (MWK) of the Lower Saxony State.
\end{acknowledgments}

\bibliography{bibliography}
\end{document}